\documentclass{aa}  

\usepackage{graphicx}
\usepackage{txfonts}
\usepackage{subfigure}
\usepackage{color}
\usepackage[none]{hyphenat}

\begin{document} 

   \title{The full spectral radiative properties of Proxima Centauri}

   \author{Ignasi Ribas\inst{1}
          \and
          Michael D. Gregg\inst{2}
          \and
          Tabetha S. Boyajian\inst{3}
          \and
          Emeline Bolmont\inst{4}
          }

   \institute{Institut de Ci\`encies de l'Espai (IEEC-CSIC), C/Can Magrans, 
s/n, Campus UAB, 08193 Bellaterra, Spain\\
              \email{iribas@ice.cat}
         \and
Department of Physics, University of California, Davis, One Shields Avenue, Davis, CA 
95616, USA
         \and
Department of Physics \& Astronomy, Louisiana State University, Baton Rouge, LA 70803, USA
         \and
Laboratoire AIM Paris-Saclay, CEA/Irfu Universit\'e Paris-Diderot CNRS/INSU, 91191 
Gif-sur-Yvette, France
             }
   \date{Received; accepted}

\abstract
{The discovery of Proxima b, a terrestrial temperate planet, presents the opportunity 
of studying a potentially habitable world in optimal conditions. A key aspect to model 
its habitability is to understand the radiation environment of the planet in the 
full spectral domain.}
{We characterize the X-rays to mid-IR radiative properties of Proxima with 
the goal of providing the top-of-atmosphere fluxes on the planet. We also aim at 
constraining the fundamental properties of the star, namely its mass, radius, effective 
temperature and luminosity.}
{We employ observations from a large number of facilities and make use of different
methodologies to piece together the full spectral energy distribution of Proxima. In the 
high-energy domain, we pay particular attention to the contribution by rotational 
modulation, activity cycle, and flares so that the data provided are representative of 
the overall radiation dose received by the atmosphere of the planet.}
{We present the full spectrum of Proxima covering 0.7 to 30000 nm. The integration of 
the data shows that the top-of-atmosphere average XUV irradiance on Proxima b is 
0.293~W~m$^{-2}$, i.e., nearly 60 times higher than Earth, and that the total 
irradiance is $877\pm44$~W~m$^{-2}$, or 
$64\pm3$\% of the solar constant but with a significantly redder spectrum. 
We also provide laws for the XUV evolution of Proxima corresponding to two scenarios, one with a 
constant XUV-to-bolometric luminosity value throughout its history and another one in which
Proxima left the saturation phase at an age of about 1.6 Gyr and is now in a power-law regime.
Regarding the fundamental properties of Proxima, we find $M=0.120\pm0.003$~M$_{\odot}$, 
$R=0.146\pm0.007$~R$_{\odot}$, $T_{\rm eff}=2980\pm80$~K, and $L=0.00151\pm0.00008$~L$_{\odot}$. 
In addition, our analysis reveals a $\sim$20\% excess in the 3--30~$\mu$m flux of the star 
that is best interpreted as arising from warm dust in the system.} 
{The data provided here should be useful to further investigate the current atmospheric
properties of Proxima b as well as its past history, with the overall aim of firmly
establishing the habitability of the planet.} 


   \maketitle
%

\section{Introduction}

The discovery of a terrestrial planet candidate around the nearest star to the Sun,
Proxima Centauri (hereafter Proxima), was reported by \citet{AngladaEscudeetal2016}
and has opened the door to investigating the properties of a potentially habitable planet 
from nearest possible vantage point. The detailed studies of \citet{Rea16} and 
\citet{Tea16} show that Proxima b is likely to have undergone substantial loss of 
volatiles, including water, in particular during the first $\sim$100--200 Myr, when 
it could have been in a runaway phase prior to entering the habitable zone. Volatile 
loss processes once inside the habitable zone could have also been at work. The 
calculations are highly uncertain \citep[cf.][]{Bea16} and reasonable doubt exists 
as to whether the modelling schemes currently used are adequate. There are numerous 
examples in the Solar System that would contradict the hypothesis of substantial 
volatile losses in the early stages of its evolution in spite of the Sun being a 
strong source of high-energy radiation \citep{M12}. The studies carried out thus far 
conclude that Proxima b is a viable habitable planet candidate because the presence 
of surface liquid water cannot be ruled out, as the initial amount of water is uncertain
and the efficiency of volatile loss processes is poorly known.

A key ingredient for understanding the evolution and current state of the atmosphere
of Proxima b is a proper description of the high-energy irradiation. Today, the 
flux that Proxima b receives in the XUV domain (X-rays to EUV) is stronger than that 
received by the Earth by over an order of magnitude and the level of irradiation was 
probably even stronger in the past. The situation is likely to be quite different in 
the UV range as Proxima has a significantly lower photospheric temperature than the 
Sun and therefore a redder emission distribution. UV irradiation has an impact on 
photolysis processes, as photoabsorption cross sections of abundant molecules peak in 
the 100--300 nm range \citep{H71}, and is also of biological interest \citep{RS16}. 
Therefore, the high-energy budget from the X-rays to the UV is important for many aspects 
related to the study of Proxima b, including understanding its atmospheric physical 
properties, its photochemistry, and even to the first attempts to constrain a putative 
biosphere on its surface. The optical and IR irradiation, on the other hand, is the main
contributor to the overall energy budget, thus determining the surface temperature of 
the planet and, ultimately, its habitability.

\citet{Rea16} obtained a rough XUV spectrum of Proxima and also discussed possible
XUV evolution laws. Here we generalize this study by providing better estimates of the 
radiation environment of Proxima b and extending the analysis to the full spectral domain
(X-rays to mid-IR). In Sect. \ref{SED} we combine observations over a wide wavelength 
range to deduce the spectral energy distribution (SED) of Proxima that is representative of 
the average radiation dose. As a consequence of this analysis, we identify a conspicuous 
IR excess, possibly due to dust in the Proxima system, which is discussed in Sect. 
\ref{IRexcess}. Also important to understand the climate of Proxima b is a good 
determination of the basic physical properties of its stellar host, namely its mass, 
radius, effective temperature and bolometric luminosity. In Sect. \ref{Lbol} we use 
all available observational constraints to provide the best estimate of such properties. 
In Sect. \ref{XUV} we address the issue of the XUV evolution law and propose two 
relationships that take into account the pre-main sequence evolution of Proxima. We 
also perform a new calculation of water loss during the early stages of the evolution 
of Proxima b and we compare the results with our earlier estimates in \citet{Rea16}. 
Finally, the conclusions of our work are given in Sect. \ref{concl}.

\section{Spectral energy distribution} \label{SED}

The aim of the study is to provide the full energy distribution at the top of the 
atmosphere of Proxima b by characterizing the electromagnetic spectrum of the host 
star as accurately as possible. This necessarily implies making use of a number of 
different facilities and also employing theoretical estimates for those wavelength 
intervals that do not have observations. Some of the datasets that we consider were 
already discussed in \citet{Rea16} and we just give a short description and additional 
relevant details. We have also improved the methodology in the case of the FUV range
(Sect. \ref{FUSE}) and this leads to a total integrated XUV flux value that differs 
by a few per cent from that presented by \citet{Rea16}. A summary of the wavelength
intervals considered and the datasets used is provided in Table \ref{tab:facilities},
and the full details are discussed in the sections below.

\begin{table*}
    \caption{List of facilities, instruments and methods employed to determine the
full spectral energy distribution of Proxima.}
    \centering
    \scriptsize
    \begin{tabular}{lllcl}
    \hline
    \hline
$\lambda$ range (nm) & Facility/Instrument & Dataset & Section & Method\\
\hline
0.7--3.8  & XMM-Newton/RGS & 0551120201, 0551120301, 0551120401 & \ref{XMM} & Combined spectrum, energetic flare correction\\
3.8--10   & ROSAT/PSPC & RP200502N00, RP200502A01, RP200502A02, RP200502A03 & \ref{ROSAT} & Plasma fit, combined spectrum, energetic flare correction\\
10--40  & EUVE  & proxima\_cen\_\_9305211911N  & \ref{EUVE} & Spectrum, energetic flare correction \\
40--92  & Model &    & \ref{Lyman_cont} & Scaled from H Lyman $\alpha$ \\
92-117  & FUSE  & D1220101000 & \ref{FUSE} & Spectrum, geocoronal correction \\
117--121.4 \& & HST/STIS E140M &  O5EO01010, O5EO01020, O5EO01030,
O5EO01040 & \ref{STISE} & Combined spectrum \\
121.7--170 &&O5EO02010, O5EO02020, O5EO02040 &&\\
121.4--121.7    & HST/STIS E140M & O5EO01010, O5EO01020, O5EO01030,
O5EO01040   & \ref{STISE} & Fit to H Lyman $\alpha$ wings to correct for ISM absorption \\
170--1000 & HST/STIS G230LB, & OCR7QQANQ, OCR7QQAOQ, OCR7QQARQ, OCR7QQASQ, & \ref{STISFOS} & Spectrum \\
 & G430L,G750L & OCR7QQAMQ, OCR7QQAPQ, OCR7QQAQQ & &  \\
1000--30000 & Model & IR photometry (Table \ref{tab:IRphot}) & \ref{Model} & Fit using BT-Settl models \\
\hline
\hline
    \end{tabular}
    \label{tab:facilities}
\end{table*}

One of the complications associated with the determination of the flux emitted by Proxima
is the effect of stellar flares. Flare events can significantly increase the flux with a
relative contribution that is stronger at shorter wavelengths. In the present study we 
estimate the mean XUV flux over a relatively extended timescale in an attempt 
to measure the overall dose on the planetary atmosphere, including the flare contribution. 
Our strategy is, thus, to consider long integration times to ensure proper averaging of the 
flare events with the quiescent flux. We apply a further correction to account for the 
contribution of large (infrequent) flares, and this correction is $\sim$10--25\%, depending 
on wavelength, with larger corrections for shorter wavelengths. The basic scheme is the 
same as in \citet{Rea16}, and the actual details are discussed for each wavelength interval 
below. Of course, future detailed multiwavelength studies of Proxima flares can provide 
much better constraints.

\subsection{X-rays: XMM} \label{XMM}

\begin{figure}
\centering
\includegraphics[width=\linewidth]{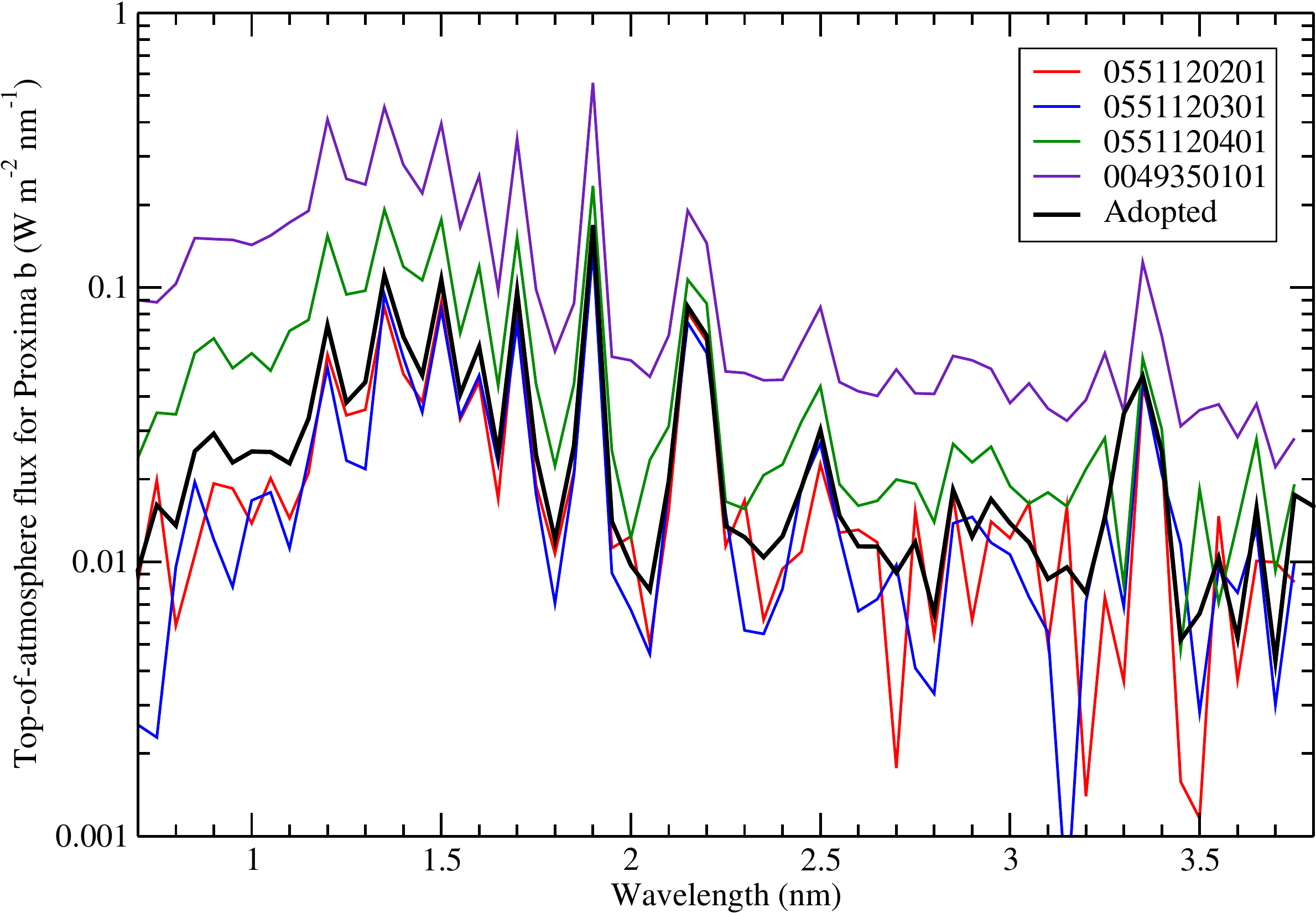}
\caption{Comparison of the top-of-atmosphere fluxes for Proxima b corresponding
to 4 XMM-Newton datasets (in color). The spectrum with higher fluxes (0049350101) corresponds 
to an observation that includes a strong flare event. Our adopted representative mean 
flux is depicted in black, and has been corrected to the mean point of the activity cycle and 
considers contributions from flares.}
\label{fig:xmm}
\end{figure}

In the 0.7 to 3.8 nm range we used four XMM-Newton observations with IDs 0049350101, 
0551120201, 0551120301, and 0551120401. The first dataset, with a duration of 67 ks, 
was studied by \citet{Guedeletal2004} and contains a very strong flare with a 
total energy of $\approx2\times10^{32}$ erg. The other three datasets (adding to 
a total of 88 ks), were studied by \citet{Fuhrmeisteretal2011}, and include several
flares, the strongest of which has an energy of about $2\times10^{31}$ erg. As in 
\citet{Rea16} we adopt the total spectrum corresponding to the combined 88-ks datasets 
and an additional energetic flare correction corresponding to a flux multiplicative 
scaling factor of 1.25. The comparison of the four individual observations with our 
adopted spectrum is shown in Fig. \ref{fig:xmm}. The observation with the strong flare 
has fluxes that are 3--5 times higher than our average representative spectrum. 

X-ray observations of Proxima were also obtained with 
other facilities, namely the Chandra observatory and the Swift mission. Chandra is optimized 
for high spatial and spectral resolution, which is not relevant to the determination of the 
SED of Proxima, and its flux calibration has larger uncertainty than that of XMM-Newton 
\citep{Pea17}. As for Swift, both its sensitivity and spectral resolution are significantly 
below that of XMM-Newton. Thus, including Chandra and Swift observations in our analysis 
would not contribute significantly to the quality of the derived SED but instead add 
complexity and potential for systematic errors. For these reasons, we prefer to base our 
hard X-ray SED solely on XMM-Newton data.

A detailed analysis of Proxima X-ray observations obtained by Swift and other facilities 
was recently published by \citet{Wea17} and they find good consistency between the different 
integrated X-ray flux measurements. Furthermore, the authors present evidence of a $\sim$7-yr 
activity cycle with an amplitude of $L_{\rm X}^{\rm max}/L_{\rm X}^{\rm min} \approx 1.5$ 
and note that the XMM observations (which are the same we use) correspond to X-ray cycle 
maxima (years 2001 and 2009). This implies that a correction should be made to refer them 
to the cycle average. We did so by adopting a multiplicative factor of 0.83 applied to 
the fluxes to yield our final values.

\subsection{X-rays: ROSAT} \label{ROSAT}

ROSAT observations were used in the wavelength range from 3.8 to 10 nm. Four suitable 
datasets are available from the ROSAT archive, with IDs RP200502N00, RP200502A01, 
RP200502A02, and RP200502A03. Their integration times were 3.8, 7.9, 20.3, and 3.8 ks, and 
the observation dates 1992.3, 1993.2, 1993.7, and 1994.2, respectively. The analysis 
procedure is explained in \citet{Rea16}. We calculated the average spectrum by using the 
integration time as the weight factor, and this should correspond a mean date of 1993.5, 
which is quite close to the midpoint of the activity cycle according to \citet{Wea17}.
Comparison with the overlapping wavelength region with the XMM data indicates very good 
mutual agreement. A multiplicative scaling factor of 1.25 was further applied to include 
the energetic flare correction also in accordance with the procedure followed for XMM.

\subsection{EUV: EUVE} \label{EUVE}

For the extreme-UV range, covering from 10 to 40 nm, we used the EUVE spectrum 
available from the mission archive with Data ID proxima\_cen\_\_9305211911N, 
corresponding to an integration time of 77 ks and observation date 1993.5. The details 
of this observation are given in \citet{Rea16} and \citet{Linskyetal2014}. We corrected 
this spectrum using a multiplicative scaling factor 1.25 to account for the average flux 
contribution coming from energetic flares. No activity cycle correction was necessary
because the observation is close to the actual mid point \citep{Wea17}.

\subsection{EUV: Lyman continuum} \label{Lyman_cont}

The interval between 40 and 92~nm (Lyman limit) cannot be observed from Earth 
due to the very strong interstellar medium absorption, even for a star as nearby 
as Proxima. To estimate the flux in this wavelength range we make use of the 
theoretical calculations presented by \citet{Linskyetal2014}. We adopt the model
corresponding to intermediate activity (1303) because it best reproduces the
H Ly $\alpha$ flux at the stellar surface (see Sect. \ref{STISE}). We consider the 
wavelength intervals 40--50 nm, 50--60 nm, 60--70 nm, 70--80 nm, and 80--91 nm, 
and the resulting ratios of the fluxes to the integrated H Ly$\alpha$ flux are 
0.01, 0.04, 0.03, 0.05, and 0.12, respectively. In our combined spectrum we consider 
these wavelength bins, yielding the appropriate integrated flux values. Note that the 
flux in this interval had been underestimated by about a factor of 2 in our previous 
calculations in \citet{Rea16}.

\subsection{FUV: FUSE} \label{FUSE}

Data from FUSE were used to obtain the flux in part of the far-UV range, from 92 to 
117 nm. We employed the spectrum with Data ID D1220101000 with a total integration
time of 45 ks and observation date 2003.3 (another FUSE dataset exists, namely 
P1860701000, but it has much shorter duration - 6 ks - and correspondingly lower 
signal-to-noise ratio). All obvious spectral regions with geocoronal emission were 
removed and only the wavelength intervals with stellar features \citep[using the 
spectrum in][as a reference]{Redfieldetal2002} were kept. The actual intervals are: 
97.4--98 nm, 99.1--101.1 nm, 103.0--103.4 nm, 103.7--103.8 nm, 110.9--113 nm. These 
wavelength ranges include most of the features from stellar origin (notably three strong 
lines corresponding to C~{\sc iii} and O~{\sc vi}, which account for 80\% of the 
92--117 nm flux except for the H Lyman series) and no geocoronal contamination. These 
intervals are missing the flux from the H Lyman series from H Ly $\beta$ to the H 
Lyman limit and this needs to be considered. 

We calculated the ratios between the different H Lyman features using an 
intermediate activity model (1303) from \citet{Linskyetal2014}. The values are 
shown in Table \ref{tab:Lyman}. To produce a spectrum, we assumed the line profile 
from the H Ly $\alpha$ feature (see Sect. \ref{STISE}). For each of the H Lyman series 
lines we scaled the width to match the typical width of the stellar features 
(C~{\sc iii} and O~{\sc vi}) and also the height of the emission to match the 
integrated flux. The results that we obtain are consistent with those presented by 
\citet{Guinanetal2003} and \citet{Ribasetal2005} for a Sun-like star with similar 
scaled H Lyman $\alpha$ flux.

\begin{table}
    \caption{Ratios between the different H Lyman features using an 
intermediate activity model (1303) from \citet{Linskyetal2014}.}
    \centering
    \begin{tabular}{lrc}
    \hline
    \hline
Feature & Wavelength (nm) & Ratio to H Ly $\alpha$\\
\hline
H Ly $\beta$   & 102.57 & 0.0195 \\
H Ly $\gamma$  &  97.25 & 0.0089 \\
H Ly $\delta$  &  94.97 & 0.0057 \\
H Ly $\epsilon$&  93.78 & 0.0037 \\
H Ly 7         &  93.08 & 0.0025 \\
H Ly 8         &  92.62 & 0.0017 \\
H Ly 9         &  92.31 & 0.0012 \\
H Ly 10        &  92.10 & 0.0007 \\
H Ly 11+rest   &  91.2--91.9 & 0.0045\\
\hline
\hline
    \end{tabular}
    \label{tab:Lyman}
\end{table}

\citet{Christianetal2004} found 3 flare events in the FUSE dataset that
we employed, which produce an increase of up to one order of magnitude in the 
instantaneous flux. The combined effect of such flares is about 20--30\% relative 
to the quiescent emission, which appears to be reasonable given our X-ray estimates 
below. Also, the observation date is close to the mid point of the activity cycle
\citep{Wea17}. Thus, no further corrections were applied.

\subsection{FUV: HST/STIS E140M} \label{STISE}

A high-quality spectrum from the StarCAT catalog \citep{Ayres2010} obtained with 
the HST Space Telescope Imaging Spectrograph \citep[STIS;][]{Wea98} was used to 
measure the fluxes between 117 and 170~nm  (except for H Ly$\alpha$). The spectrum 
was produced by co-adding a number of individual observations corresponding to HST 
datasets O5EO01010, O5EO01020, O5EO01030, O5EO01040, O5EO02010, O5EO02020, and 
O5EO02040 and with a total integration time of 35.7 ks and observation date
2000.4. A flare analysis of these individual datasets was carried out by 
\citet{LoydFrance2014}, who identified a number of flare events in the stronger 
emission lines. These flares contribute some 25--40\% of the integrated flux (Loyd, 
priv. comm.) and thus represent similar values to those found in X-rays. In addition, 
as before, the date of the observations is nearly at the mid point of the activity 
cycle \citep{Wea17} and no further corrections were made. The intrinsic line profile 
of the H Ly$\alpha$ feature that we adopt was calculated from the same base spectrum 
by \citet{Woodetal2005}. The relative flare contribution corrected for ISM absorption 
is estimated to be of $\sim$10\% (Loyd, priv. comm.). 

\subsection{UV to NIR: HST/STIS \& HST/FOS} \label{STISFOS}

Proxima was observed with HST/STIS on 24 April, 2015 as part of the Cycle~22 incarnation 
of the Next Generation Spectral Library (NGSL). The specific dataset references are 
listed in Table \ref{tab:facilities}. The NGSL is an HST/STIS snapshot 
program which has compiled a spectral library of 570 representative spectral 
type stars for use in spectral synthesis of galaxies and other composite stellar 
systems. The spectra are obtained using the three low dispersion CCD modes of 
STIS, G230LB, G430L, and G750L, covering $\lambda\lambda$170--1020~nm at
a resolution of about 1000. For Proxima, the exposure times were $2\times600$~s, 
$2\times30$~s, and 30~s for the three gratings.

The G230LB and G430L spectra were obtained through the $0\farcs2$ E1 aperture, 
located near one edge of the CCD to reduce charge transfer losses during readout.  
The G750L spectrum, also observed through a $0\farcs2$ slit, was obtained at the 
regular long slit center near the middle of the CCD. This was in order to take 
advantage of the very narrow $0\farcs09$ slit during contemporaneous CCD fringe 
flat calibration exposures to improve the removal of the considerable (10--15\%) 
fringing above 700~nm in the G750L data.

To save valuable on-target time during the HST snapshots, no contemporaneous 
wavelength calibrations ({\sc wavecals}) are carried out during NGSL observations.  
Instead, a generic wavelength calibration is supplied in the pipeline download 
of the data, and a linear zeropoint pixel shift is determined either from inspection 
or cross-correlation of a preliminary 1D extraction of the source with a velocity 
template spectrum.  This pixel shift is then inserted into the FITS header of the 2D 
STIS data; subsequent extraction of 1D spectra using the task {\sc x1d} in the 
{\sc stsdas} package of {\sc iraf} takes out the first-order grating setting difference 
between the actual observation and the generic wavelength solution, typically
3--5 pixels.

The 1D spectra were extracted in {\sc iraf/pyraf} using the {\sc x1d} task. During 
extraction, the {\sc x1d} task also applies charge transfer inefficiency corrections, 
corrects for slit losses in the $0\farcs2$ slit, and applies an overall flux calibration 
to units of $F_{\lambda}$. The G750L spectrum was defringed using the contemporaneous 
fringe flat obtained through the narrower slit which mimics a point source on the detector 
better than obtaining a flat through the $0\farcs2$ slit.

The G230LB mode of STIS suffers from contamination by scattered zero-order light from 
all wavelengths to which the detector is sensitive. This is corrected for using the 
procedure developed by \citet{LH08}.  Briefly, the initial combined flux calibrated 
spectrum is run in reverse through the G230LB sensitivity function covering all 
wavelengths to produce a best estimate of the G230LB counts over the entire optical 
range. From this, a pixel-by-pixel correction is calculated from a simple
wavelength-dependent function dependent on the total counts in the overall 
computed spectrum.  This correction is then subtracted pixel-by-pixel from the G230LB 
counts spectrum to correct for the red light contamination of the UV spectrum, and 
then the G230LB counts spectrum is again flux calibrated.

Inspection of the modest wavelength overlap ($\sim$200~nm) between the three low 
dispersion spectra shows that the absolute calibrations of three individual gratings 
agree to better than 2--3\% for the observation of Proxima. The three individual grating
spectra were combined into a single spectrum using the {\sc scombine} and {\sc dispcor} 
tasks in {\sc iraf}. The final spectrum covers 170 to 1020~nm with a constant sampling 
0.2~nm per pixel.

\subsubsection{Comparison to ground-based photometry}

Standard $UBVRI$ photometry of Proxima was collected from the literature. Measurements 
from different sources are provided in Table \ref{tab:optphot}. In the case of the $U$ 
band, one of the measurements is very discrepant from the other two. It is possible
that the value from \citet{Jea14}, which is brighter by $0.3$ mag and corresponds 
to a single epoch, was affected by a flare. In contrast, the photometric measurements of
\citet{Fea72} correspond to the average of several observations taken outside of 
flare activity and thus we adopt the $U$-band magnitude from this study. The quoted 
uncertainty is 0.05 mag. For the $BVRI$ bands we adopt the photometry from \citet{R82}, 
which is the average of 24 individual measurements. The quoted uncertainty is
0.028 mag, although it is not certain to which band this value corresponds. 

\begin{table}
\scriptsize
\caption{Optical $UBVRI$ photometry of Proxima. Magnitudes are in the 
Johnson-Cousins system unless otherwise noted.}
\begin{tabular}{llllll}
\hline
\hline
\multicolumn{1}{c}{$U$} & \multicolumn{1}{c}{$B$} & \multicolumn{1}{c}{$V$} 
& \multicolumn{1}{c}{$R$} & \multicolumn{1}{c}{$I$} & Ref \\
\hline
14.55 & 13.12  & 11.22 & 8.970\tablefootmark{a}& 7.310\tablefootmark{a} & \citet{MH76}\\
14.56 & 13.02  & 11.04 & 8.68\tablefootmark{a} & 6.42\tablefootmark{a}  & \citet{Fea72}\\
      & 12.988 & 11.11 & 9.429    & 7.442     & \citet{R82}\\
14.21 & 12.95  & 11.13 & 9.45     & 7.41      & \citet{Jea14}\\
      & 13.02  & 11.05 &          &           & \citet{GJ15}\\ 
      &        & 11.05 & 9.43     & 7.43      & \citet{B91}\\
\hline
14.56 & 12.988 & 11.11 & 9.429    & 7.442     & Adopted \\
\hline
14.492 & 13.000 & 11.147 &  9.399   &  7.374    & STIS synthetic \\
14.491 & 12.970 & 11.083 &  9.380   &  7.384    & STIS+FOS synthetic \\
\hline
\end{tabular}
\tablefoot{
\tablefoottext{a}{Not in the Cousins photometric system. No reliable transformation to 
Cousins for such red object is available.}}
\label{tab:optphot}
\end{table}

To compare the final STIS spectrum with the ground-based photometry, we calculated 
synthetic Johnson/Cousins photometric indices by convolving the STIS spectrum with 
$UBVRI$ bandpasses from \citet{BM12}, with zero points calibrated via the STIS\_008 
Vega spectrum from the CALSPEC Calibration database at 
\url{http://www.stsci.edu/hst/observatory/crds/calspec.html}. The calculated magnitudes 
are listed in Table \ref{tab:optphot}. All bands agree with the adopted, best-reliable 
ground-based photometry of Proxima within 3--6\%. 

\subsubsection{Comparison to HST/FOS}

\begin{figure}
\centering
\includegraphics[width=\columnwidth]{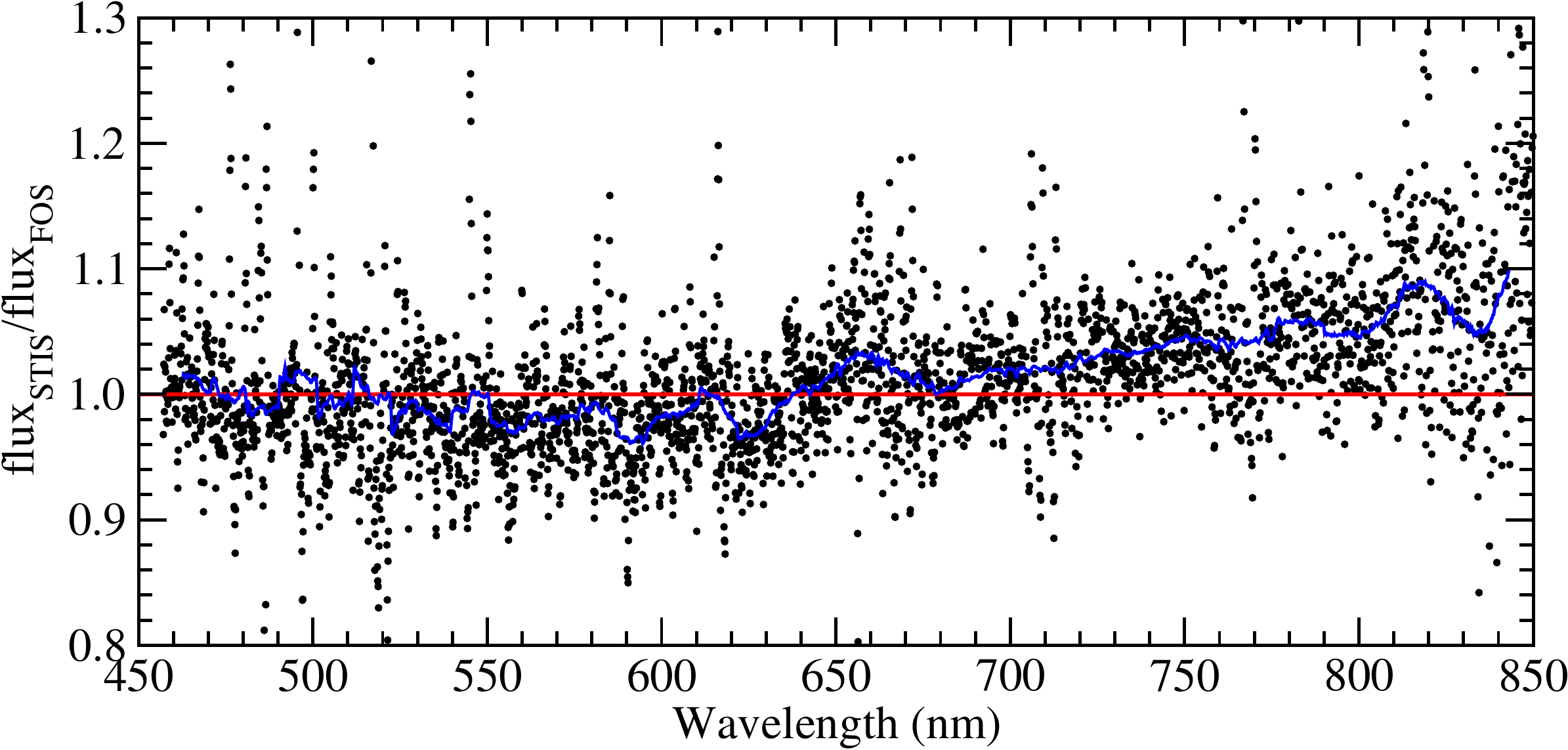}
\caption{Ratio of STIS to FOS fluxes shown as black circles for each wavelength bin. The 
blue line shows a 100-point running average.}
\label{fig:STISFOSratio}
\end{figure}

HST observed Proxima with the Faint Object Spectrograph (FOS) on 1 July, 1996 through 
the $1\farcs0$ aperture for 430 s with the G570H grating (dataset Y2WY0305T) and 280 s 
with the G780 grating (dataset Y2WY0705T). The spectrum covers $\lambda\lambda$450--850~nm 
with a resolution of 0.09~nm and we compare the STIS and combined FOS spectra in Fig. 
\ref{fig:STISFOSratio}. There is general agreement at the $\sim$5\% level and thus
the STIS spectrum agrees with the FOS data of Proxima at a similar level to the 
broad-band photometry.

\subsubsection{Final spectrum}

The FOS spectrum covers a subrange of the STIS spectrum and it does so at a higher 
spectral resolution. One could thus consider adopting a final spectrum composed of 
three wavelength intervals: 170--460 nm (STIS), 460--840 nm (FOS), 840--1000 nm
(STIS). We have calculated the comparison with broad-band photometric measurements 
and this is shown in the last row of Table \ref{tab:optphot}. As expected, the 
differences are rather minor with the all-STIS spectrum. Also, for most applications 
requiring irradiance measurements, the increased resolution in the central part of 
the optical wavelength range is of little use. Given these considerations, the 
results of the comparison between the STIS and FOS fluxes, and the interest of 
preserving homogeneity, we decided to adopt the full wavelength coverage from STIS
as a fair representation of the spectral energy distribution of Proxima over the 
wavelength region of comparison.  

As occurs at high energies, Proxima is also known to experience flux variations in 
the optical due to the presence of surface inhomogeneities 
\citep{AngladaEscudeetal2016,Wea17}. Photometric monitoring of Proxima shows that the 
peak-to-peak variability with respect to the mean is of the order of 5\% in the $B$ 
band and 2\% in the $V$ band over timescales of months, and can be attributed 
to rotational modulation. This provides a viable explanation for the 3--6\% difference in 
the results of the comparisons between different measurements. The analysis of \citet{Dea16} 
using $MOST$ satellite observations covering roughly 430 to 760 nm and taken over a time 
period of nearly 38 days reveal frequent white-light flares. There are 5--8 measurable 
flares per day with a typical duration of $\sim$1 hour. However, the average flux 
contribution from flares in this wavelength range to the quiescent flux is only 2.6\% 
(Davenport, priv. comm.). This relatively small effect, less than the typical uncertainty 
of the absolute flux calibration, suggests that flare correction to optical (and IR) 
spectrophotometric and photometric observations is not necessary. The flux values for 
Proxima that we provide should be representative of the average flux to better than 5\%.

\subsection{IR: Model spectrum} \label{Model}

As we have shown, spectrophotometric observations that can be calibrated to yield
physical fluxes are available for most wavelength regions up to about 1~$\mu$m. Beyond this
wavelength value, the measurements are in the form of broad-band magnitudes or fluxes. We 
performed a search in the literature for flux measurements of Proxima. An important
source of measurements is the catalog of \citet{Gea99}, and we complemented it with
subsequent references. A summary of the photometry is given in Table \ref{tab:IRphot}. 
In view of the uncertainties and absolute calibration of the photometric systems, we decided
to adopt two independent photometric datasets, namely the 2MASS $JHK$ photometry (although
the 2MASS $K$ band measurement has a flag indicating poor quality) and the \citet{MH76} $JHKL$
photometry, which has two epochs and is given in a well-calibrated standard system. No
$M$-band photometry was used in view of the large uncertainty. The magnitudes were 
transformed into physical flux units using the zero-point calibrations in \citet{Coea03} 
for the 2MASS system and \citet{Bea98} for the photometry from \citet{MH76}. The fluxes 
are given in Table \ref{tab:IRflux}. We additionally considered the recent revision of 
the zero points for the NIR magnitudes by \citet{Mv15} but the results are very similar. 

\begin{table*}
\small
\centering
\caption{Infrared $JHKLM$ photometry of Proxima. The adopted values are highlighted
in bold face.}
\begin{tabular}{llllll}
\hline
\hline
\multicolumn{1}{c}{$J$} & \multicolumn{1}{c}{$H$} & \multicolumn{1}{c}{$K$} 
& \multicolumn{1}{c}{$L$} & \multicolumn{1}{c}{$M$} & Ref \\
\hline
5.34            & 4.71            & 4.36                &             &             & \citet{B91} (Glass system) \\
                & 4.73$\pm$0.05   & 4.40$\pm$0.05   & 4.17$\pm$0.01   &             & \citet{Fea72} (SAO system) \\
                &                 & 4.6$\pm$ 0.1        & 4.1$\pm$0.1 & 4.0$\pm$0.2 & \citet{Fea72} (Minnesota system) \\
5.39$\pm$0.03   & 4.74$\pm$0.02   & 4.38$\pm$0.02       & 4.15        &             &  \citet{AP91} (CTIO system) \\
\textbf{5.357$\pm$0.023} & \textbf{4.835$\pm$0.057} & \textbf{4.384$\pm$0.033\tablefootmark{a}} &             &             & \citet{Cea03} (2MASS) \\
                & 4.73$\pm$0.1    & 4.40$\pm$0.1        &             &             & \citet{V74} (Johnson system) \\
\textbf{5.330$\pm$0.020} & \textbf{4.725$\pm$0.020} & \textbf{4.365$\pm$0.028}     &\textbf{4.04$\pm$0.06}&             &          \citet{MH76} (Glass system) \\
\hline
\hline
\end{tabular}
\tablefoot{
\tablefoottext{a}{Saturated, flag E.}}
\label{tab:IRphot}
\end{table*}

\begin{table*}
\centering
\caption{Infrared fluxes of Proxima.}
\begin{tabular}{llcccccl}
\hline
\hline
Band  & $\lambda_{\rm eff}$ & Magnitude & Error & Flux & Error & Flux (STIS\_008) & Passband\\
      & ($\mu$m)            &           &       & 
      {\scriptsize (erg s$^{-1}$ cm$^{-2}$ \AA$^{-1}$)} & 
      {\scriptsize (erg s$^{-1}$ cm$^{-2}$ \AA$^{-1}$)} & 
      {\scriptsize (erg s$^{-1}$ cm$^{-2}$ \AA$^{-1}$)} & source \\
\hline
G750L       &1.00&      &       & $2.37\times10^{-12}$ & $4.7\times10^{-14}$ &                & Square (FWHM=40 nm)\\
$J$ (2MASS) &1.24& 5.357& 0.023 & $2.25\times10^{-12}$ & $5.6\times10^{-14}$ &  $2.25\times10^{-12}$ & \citet{Cea03}\\
$H$ (2MASS) &1.65& 4.835& 0.057 & $1.32\times10^{-12}$ & $7.9\times10^{-14}$ & $1.33\times10^{-12}$ & \citet{Cea03}\\
$K$ (2MASS) &2.16& 4.384& 0.033 & $7.55\times10^{-13}$ & $2.7\times10^{-14}$ & $7.55\times10^{-13}$ & \citet{Cea03}\\
$J$ (MH76)  &1.24& 5.330& 0.020 & $2.32\times10^{-12}$ & $4.7\times10^{-14}$ & $2.34\times10^{-12}$ & \citet{Bea98}\\
$H$ (MH76)  &1.64& 4.725& 0.020 & $1.45\times10^{-12}$ & $2.9\times10^{-14}$ & $1.48\times10^{-12}$ & \citet{Bea98}\\
$K$ (MH76)  &2.19& 4.365& 0.028 & $7.11\times10^{-13}$ & $2.1\times10^{-14}$ & $7.17\times10^{-13}$ & \citet{Bea98}\\
$L$ (MH76)  &3.50& 4.04 & 0.06  & $1.71\times10^{-13}$ & $1.0\times10^{-14}$ & $1.74\times10^{-13}$ & \citet{Bea98}\\
$W1$ (AllWISE)\tablefootmark{a}&3.42& 4.207& 0.331 & $1.70\times10^{-13}$ & $5.9\times10^{-14}$ & $1.75\times10^{-13}$ & \citet{Wea10}\\
$W2$ (AllWISE)\tablefootmark{a}&4.49& 3.779& 0.131 & $7.44\times10^{-14}$ & $1.0\times10^{-14}$ & $7.59\times10^{-14}$ & \citet{Wea10}\\
$W3$ (AllWISE)\tablefootmark{a}&10.6& 3.838& 0.015 & $1.90\times10^{-15}$ & $3.8\times10^{-17}$ & $2.28\times10^{-15}$ & \citet{Wea10}\\
$W4$ (AllWISE)\tablefootmark{a}&21.8& 3.688& 0.025 & $1.70\times10^{-16}$ & $4.2\times10^{-18}$ & $1.69\times10^{-16}$ & \citet{Wea10}\\
MSX $A$     &8.38&      &       & $8.00\times10^{-15}$ & $3.3\times10^{-16}$ & & \citet{Eea99}\\
MSX $C$     &12.0&      &       & $2.02\times10^{-15}$ & $1.8\times10^{-16}$ & & \citet{Eea99}\\
IRAS\_12    &10.4&      &       & $3.18\times10^{-15}$ & $1.9\times10^{-16}$ & & \citet{IRAS94}\\
IRAS\_25    &21.8&      &       & $1.19\times10^{-16}$ & $2.3\times10^{-17}$ & & \citet{IRAS94}\\
MIPS\_24    &23.3&      &       & $1.30\times10^{-16}$ & $2.0\times10^{-19}$ & & \citet{MIPS04}\\
\hline
\hline
\end{tabular}
\tablefoot{
\tablefoottext{a}{Magnitudes from the WISE All-Sky Source catalog are $W1=4.195\pm0.086$, 
$W2=3.571\pm0.031$, $W3=3.826\pm0.035$, $W4=3.664\pm0.024$.}}
\label{tab:IRflux}
\end{table*}

Besides the photometry in the classical broad-band systems, flux measurements of Proxima
coming a number of space missions also exist, namely WISE, MSX, IRAS, and Spitzer. For the WISE 
mission \citep{Wea10} we considered both the AllWISE and the WISE All-Sky Source catalogs. 
In both cases, the $W1$ and $W2$ magnitudes are saturated (17\% to 25\% saturated pixels), 
while the $W3$ and $W4$ bands are not. The agreement for the $W1$, $W3$ and $W4$ bands for 
both catalogs is good but the $W2$ magnitudes are highly discrepant. The $W2$ magnitude 
from the WISE All-Sky Source catalog leads to an unphysical energy distribution (much 
higher flux than in all other bands). Also, the uncertainties associated to the 
saturated bands of the WISE All-Sky Source catalog seem unrealistically low. We decided to 
adopt the AllWISE measurements and uncertainties but did not consider the saturated $W1$ an 
$W2$ bands in the fits. The physical fluxes for the WISE bands were calculated using zero 
points in \citet{Jea11} and are listed in Table \ref{tab:IRflux}. Proxima is included in 
the MSX6C Infrared Point Source Catalog \citep{Eea03}. Measurements are only available in 
the so-called $A$ and $C$ bands and are given in physical units. These are included in 
Table \ref{tab:IRflux}. Proxima was also observed by the IRAS mission in two bands and 
has an entry in the IRAS catalog \citep{IRAS}, with measurements in two bands (12 $\mu$m 
and 25 $\mu$m). The fluxes are provided in physical units and are listed in Table 
\ref{tab:IRflux}. Finally, \citet{Gea07} included Proxima in their survey of the far-IR
properties of M dwarfs and obtained a flux measurement in the Spitzer/MIPS 24-$\mu$m band. 
This is listed in Table \ref{tab:IRflux}. In addition to the bands considered above, we 
also included an anchor point from the HST/STIS G750L calibrated spectrophotometry at 
a wavelength of 1 $\mu$m, taking advantage of the very precise flux calibration of HST/STIS 
and to tie in with the optical measurements. For this, we considered an {\em ad hoc} 
square passband of 40 nm in width and calculated the average flux in this wavelength 
interval. 

The fluxes in Table \ref{tab:IRflux} are quite consistent for all bands except for the 
measurements of WISE $W3$ and IRAS\_12, which correspond to nearly identical effective 
wavelengths but differ by over 50\%. While we initially employed the passband zero points 
from the literature, we explored another approach, namely the calibration of the 
fluxes using a standard spectrum. As before, we used the STIS\_008 Vega spectrum from 
the CALSPEC Calibration database. We calculated integrated fluxes for the relevant 
broad-band passbands using the definitions from the references in Table \ref{tab:IRphot} 
and used them to set the zero point of the magnitude scale. The fluxes for Proxima 
calculated in this way are also listed in Table \ref{tab:IRflux}. As expected, the 
comparison between the literature zero points and those estimated using the spectrum 
of Vega reveals little differences in most cases (less than 3\%). However, the WISE 
$W3$ band zero point is notably different (by about 20\%) when comparing both methods. 
The value that we calculate from the Vega standard spectrum leads to closer agreement 
(though still far from perfect) with the IRAS\_12 value. Given this circumstance, we 
decided to adopt the fluxes as calculated by us from the Vega spectrum for all bands 
with magnitude measurements (i.e., not for fluxes given in physical units).

\begin{table*}
\footnotesize
\centering
\caption{SED model fits to Proxima IR flux measurements. All fluxes are in units of
erg s$^{-1}$ cm$^{-2}$ \AA$^{-1}$. The difference between ``Fit 1'' and ``Fit 2'' is the number
of flux measurements used in the fitting procedure.}
\begin{tabular}{lccccccccc}
\hline
\hline
Band  & $\lambda_{\rm eff}$ & \multicolumn{4}{c}{Fit 1} & \multicolumn{4}{c}{Fit 2} \\
      & ($\mu$m)            & Flux (mod) & obs--mod & (obs--mod)/$\sigma$ & used & Flux (mod) & obs--mod & (obs--mod)/$\sigma$ & used \\
\hline
G750L          &  1.000 & $2.43\times10^{-12}$ &$-5.8\times10^{-14}$  & $-$1.2& y &$2.42\times10^{-12}$ &$-5.2\times10^{-14}$  &$-$1.1&y\\
$J$ (2MASS)    &  1.245 & $2.26\times10^{-12}$ &$-1.4\times10^{-14}$  & $-$0.2& y &$2.38\times10^{-12}$ &$-1.2\times10^{-13}$  &$-$2.2&y\\
$H$ (2MASS)    &  1.647 & $1.43\times10^{-12}$ &$-9.8\times10^{-14}$  & $-$1.2& y &$1.50\times10^{-12}$ &$-1.7\times10^{-13}$  &$-$2.1&y\\
$K$ (2MASS)    &  2.162 & $6.68\times10^{-13}$ &$ 8.7\times10^{-14}$  &  3.2  & y &$7.15\times10^{-13}$ & $4.0\times10^{-14}$  & 1.5  &y\\
$J$ (MH76)     &  1.240 & $2.26\times10^{-12}$ &$ 7.5\times10^{-14}$  &  1.6  & y &$2.37\times10^{-12}$ &$-3.4\times10^{-14}$  &$-$0.7&y\\
$H$ (MH76)     &  1.642 & $1.42\times10^{-12}$ &$ 6.4\times10^{-14}$  &  2.1  & y &$1.48\times10^{-12}$ &$-4.3\times10^{-15}$  &$-$0.1&y\\
$K$ (MH76)     &  2.195 & $6.28\times10^{-13}$ &$ 8.8\times10^{-14}$  &  4.2  & y &$6.72\times10^{-13}$ & $4.5\times10^{-14}$  & 2.1  &y\\
$L$ (MH76)     &  3.501 & $1.44\times10^{-13}$ &$ 3.0\times10^{-14}$  &  3.0  & n &$1.59\times10^{-13}$ & $1.5\times10^{-14}$  & 1.5  &y\\
$W1$ (AllWISE) &  3.419 & $1.53\times10^{-13}$ &$ 2.2\times10^{-14}$  &  0.4  & n &$1.67\times10^{-13}$ & $7.9\times10^{-15}$  & 0.1  &n\\
$W2$ (AllWISE) &  4.494 & $5.48\times10^{-14}$ &$ 2.1\times10^{-14}$  &  2.1  & n &$6.04\times10^{-14}$ & $1.5\times10^{-14}$  & 1.6  &n\\
$W3$ (AllWISE) & 10.600 & $2.05\times10^{-15}$ &$ 2.3\times10^{-16}$  &  5.0  & n &$2.33\times10^{-15}$ & $5.0\times10^{-17}$  &$-$1.1&y\\
$W4$ (AllWISE) & 21.829 & $1.46\times10^{-16}$ &$ 2.3\times10^{-17}$  &  5.4  & n &$1.67\times10^{-16}$ & $2.5\times10^{-18}$  & 0.6  &y\\
MSX $A$        &  8.382 & $6.07\times10^{-15}$ &$ 1.9\times10^{-15}$  &  5.8  & n &$6.84\times10^{-15}$ & $1.2\times10^{-15}$  & 3.5  &y\\
MSX $C$        & 12.045 & $1.46\times10^{-15}$ &$ 5.6\times10^{-16}$  &  3.1  & n &$1.67\times10^{-15}$ & $3.5\times10^{-16}$  & 1.9  &y\\
IRAS\_12       & 10.375 & $2.35\times10^{-15}$ &$ 8.3\times10^{-16}$  &  4.4  & n &$2.68\times10^{-15}$ & $5.0\times10^{-16}$  & 2.6  &y\\
IRAS\_25       & 21.790 & $1.32\times10^{-16}$ &$-1.3\times10^{-17}$  & $-$0.6& n &$1.50\times10^{-16}$ &$-3.1\times10^{-17}$  &$-$1.4&y\\
MIPS\_24       & 23.245 & $1.14\times10^{-16}$ &$ 1.6\times10^{-17}$  & 80.7 & n &$1.30\times10^{-16}$ & $2.3\times10^{-19}$  & 1.2  &y\\
\hline
\hline
\end{tabular}
\label{tab:SEDfit}
\end{table*}

Our procedure to obtain the NIR SED of Proxima is to fit all flux measurements with a 
spectrum from a theoretical model. As already mentioned in Sect. 
\ref{SED}, no specific correction for flares was made. We chose to use the BT-Settl grid 
from \citet{Baraffeetal15} in its latest version available from 
\url{https://phoenix.ens-lyon.fr/Grids/BT-Settl/CIFIST2011_2015/}. Proxima's surface
gravity and metallicity are compatible within the error bars with values of $\log g = 5.0$ 
and $[Fe/H]=0.0$ \citep{Pea16}, which are part of the BT-Settl model grid, and those 
were adopted in our SED fitting procedure. The free parameters of the fit were the 
effective temperature and the angular diameter. For the latter, however, we used 
a prior from \citet{Dea09} of $\theta = 1.011\pm0.052$ mas. From the model spectra we 
calculated integrated fluxes for all passbands using the definitions from the references
in Table \ref{tab:IRphot}. We built a $\chi^2$ statistic by comparing the model fluxes
with the observed values and adopting the usual weight proportional to $1/\sigma^2$, and 
this was minimized via the simplex algorithm as implemented by \citet{NR92}. To 
further constrain the model we doubled the weight of the anchor point at 1 $\mu$m and of 
the angular diameter measurement. 

\begin{figure}
\centering
\includegraphics[width=\columnwidth]{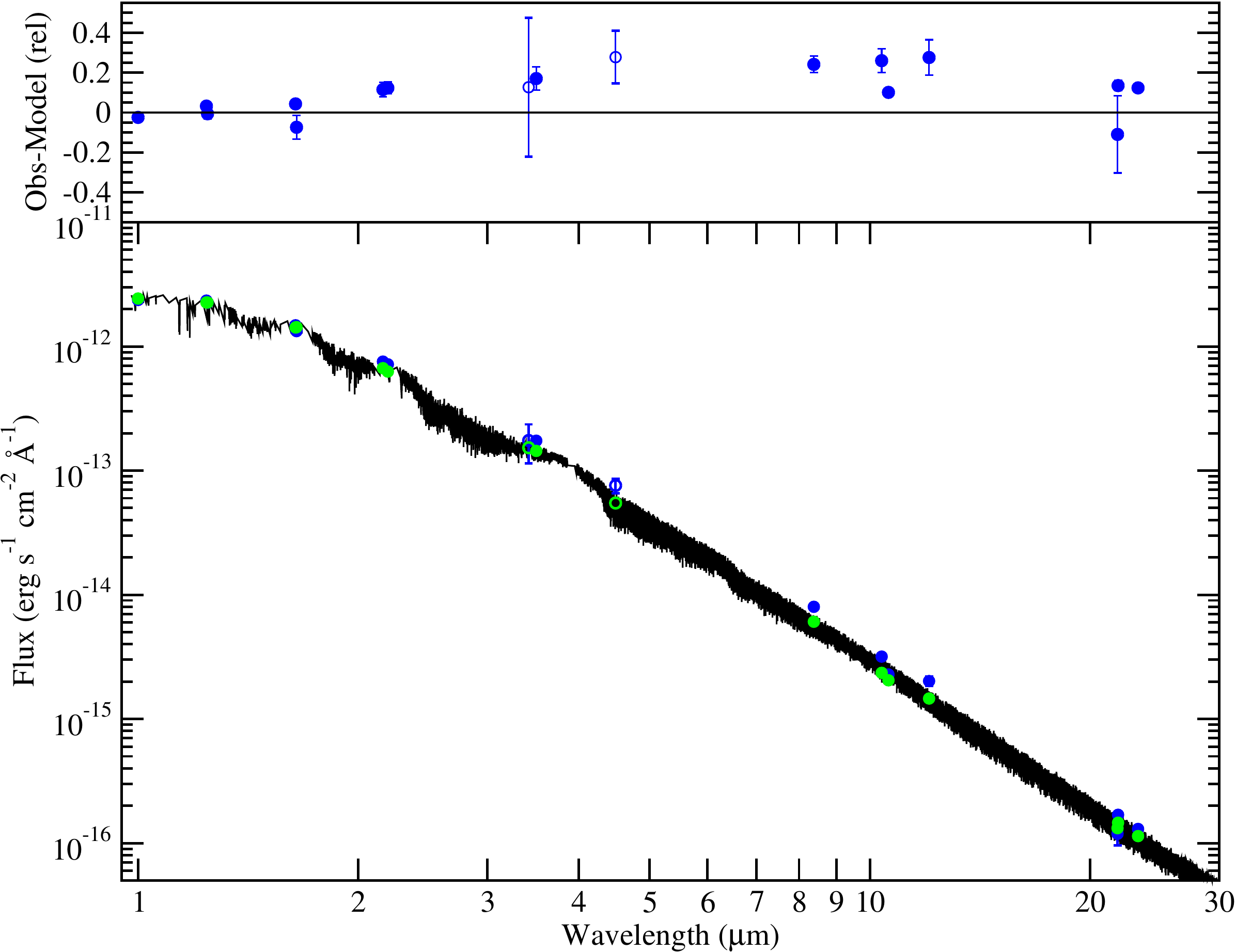}
\caption{Best-fitting model to IR fluxes considering measurements shortwards of 
3 $\mu$m. The fit residuals and parameters are given in Tables \ref{tab:SEDfit} and
\ref{tab:fitresults} under the label ``Fit 1''. The black line is the BT-Settl spectrum. The 
blue circles correspond to the observed fluxes and the green circles are the 
model-integrated fluxes. Empty symbols correspond to the WISE $W1$ and $W2$ bands, which 
are heavily saturated.}
\label{fig:IRSEDfit}
\end{figure}

Using the constraints above, the SED fit yields an effective temperature of 2870~K but 
an angular diameter that is 2.7$\sigma$ larger than the observation. Also, the residuals 
reveal a strong systematic difference between the bands roughly at either side of 3 $\mu$m. 
Such discrepancy suggests that Proxima has higher fluxes at longer wavelengths than 
expected from models. We then considered a fit only to the bands shortwards of 3 $\mu$m and 
this led to an effective temperature of 3000~K and an angular diameter within 1$\sigma$ of 
the measured value. In Table \ref{tab:SEDfit} we list the flux residuals and resulting 
parameters from two fitting scenarios. We adopted the solution that fits the bands shorter 
than 3 $\mu$m, the HST/STIS flux and $JHK$ bands, our ``Fit 1''. Figure 
\ref{fig:IRSEDfit} illustrates this fit and the comparison between models and observations. 
In addition, we ran tests by considering only measurements up to 2 $\mu$m, thereby excluding 
the $K$ band; these also show the same systematic trend and yielded very similar results. 

\begin{table}
\centering
\caption{Results from SED model fits to Proxima IR flux measurements.}
\begin{tabular}{lcc}
\hline
\hline
Parameter & Fit 1 & Fit 2 \\
\hline
$T_{\rm eff}$ (K)            & 3000  & 2870 \\
$\theta$ (mas)               & 1.042 & 1.149 \\
$\chi^2_\nu$                 & 3.48  & 8.22 \\
Flux 1--30 $\mu$m            & $2.33\times10^{-8}$  & $2.45\times10^{-8}$ \\
(erg s$^{-1}$ cm$^{-2}$ \AA$^{-1}$) (calc.)   & &\\
Radius\tablefootmark{a} (R$_{\odot}$) (calc.) & 0.146 & 0.161 \\
\hline
\hline
\end{tabular}
\tablefoot{
\tablefoottext{a}{Using the parallax measurement $\pi=0.7687\pm0.0003$ arcsec from 
\citet{Bea99}.}}
\label{tab:fitresults}
\end{table}

\subsection{Combined spectrum}

\addtocounter{table}{1}

\begin{figure*}
\centering
\includegraphics[width=\linewidth]{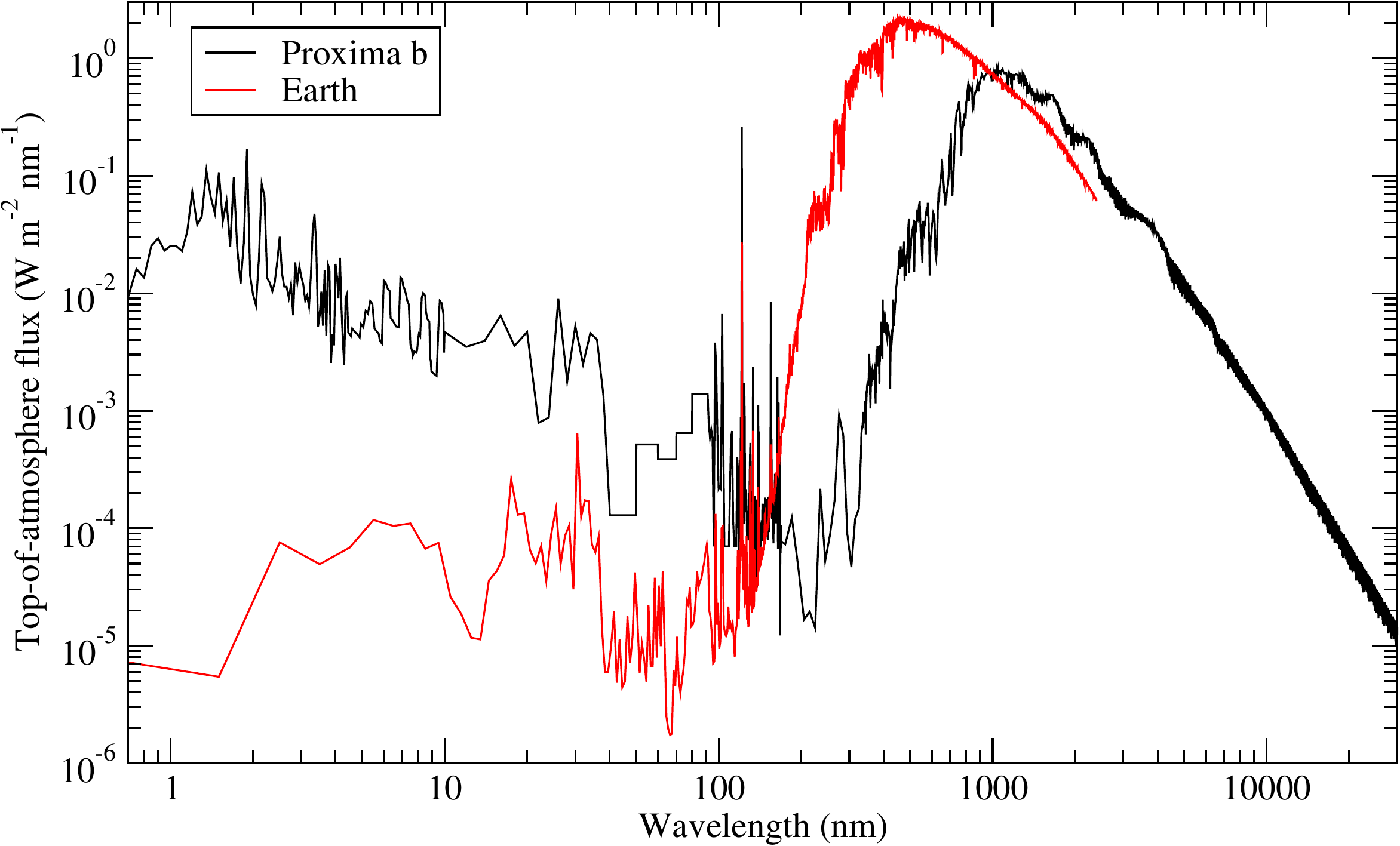}
\caption{Top-of-atmosphere full spectral irradiance received by Proxima b (black) and 
the Earth (red). An orbital distance of 0.0485 AU is assumed for Proxima b.}
\label{fig:Fullspec}
\end{figure*}

The full spectrum, covering 0.7 to 30000 nm, is provided in Table 8 and shown in Fig. 
\ref{fig:Fullspec}. Table 8 is available at the CDS and contains the following information:
Column 1 gives the wavelength in nm and Column 2 lists the top-of-atmosphere flux for 
Proxima b in units of W~m$^{-2}$~nm$^{-1}$, which are most commonly employed for planetary
atmosphere work. The spectrum as obtained by adding the data from the various sources has 
rather inhomogeneous wavelength steps and was resampled using different bin sizes for 
different wavelength intervals depending on the quality of the spectrum. We calculated 
the top-of-atmosphere flux for Proxima b by using the trigonometric distance to Proxima 
and by adopting an orbital distance of 0.0485 AU for Proxima b. Figure \ref{fig:Fullspec} 
also shows the top-of-atmosphere solar irradiance of the Earth for comparison, corresponding
to the \citet{Tea04} solar spectrum for medium solar activity. The integrated fluxes in 
various relevant intervals are listed in Table \ref{tab:XUV}. Our results show that the XUV 
flux is nearly 60 times higher than Earth's value \citep[0.0051~W~m$^{-2}$;][]{Rea16} and 
the total integrated flux is $877\pm44$ W~m$^{-2}$, or $64\pm3$\% of the solar constant 
(i.e., top-of-atmosphere solar flux received by Earth, adopting $S_{\oplus}=1361$~W~m$^{-2}$, 
\citealt{KL11}). The adopted uncertainty on the total irradiance corresponds to a relative 
error of 5\% on the Proxima flux (see below).

Proxima is variable over different timescales, most notably related to flare events 
(hours), rotational modulation (months) and activity cycle (years). In Table \ref{tab:XUV} 
we provide estimates of such variability amplitude (peak), when available, with respect to 
the mean flux value listed. Such estimates come from various literature sources that we have 
adapted to the relevant wavelength intervals. For the bolometric flux we scale the variability 
from that coming from the $V$ band. \citet{Wea17} obtain a variation of 2\% with respect to 
the average (4\% peak to peak). This, however, is not representative of the bolometric 
variability because activity-related effects are known to diminish with increasing 
wavelengths. In the case of Proxima, wavelengths around 1~$\mu$m would be a better proxy for 
flux variations of the bolometric luminosity. We have used the StarSim simulator \citep{Hea16} 
to estimate that variations of 2\% in the $V$ band correspond to about 0.5\% around 1~$\mu$m 
if we assume spots with contrasts of 300--500~K \citep{B05}. We adopt a similar scaling for 
the flare statistics obtained by \citet{Dea16}, which correspond to the MOST satellite band. 
It is interesting to point out that the variability of the total irradiance of Proxima is 
about 25 times higher than the solar value \citep[0.02\%; ][]{F12} and this could have an 
impact on the climate forcing.

It should be noted that no information on the rotational modulation and cycle 
amplitude are available for the FUV range. The only relevant data in the UV comes from 
the results of \citet{Wea17}, who find a 4\% rotational modulation and similar cycle 
variability for the Swift/UVOT $W1$ band, which has an effective wavelength of 260 nm. 
Regarding flares, \citet{W81} studied several large events in the $UBV$ bands, and found 
peak-to-quiescence flux ratios of up to 25, 4 and 1.5, respectively. We do not include these 
values in Table \ref{tab:XUV} but we note that both the rotation/cycle amplitudes and the 
flare fluxes are strongly variable with wavelength.

\begin{table*}
\centering
\caption{Top-of-atmosphere (TOA) fluxes received currently by Proxima b. Also 
provided are the peak variations with respect to the mean at various timescales related
to stellar magnetic activity.}
\begin{tabular}{lcll}
\hline
\hline
Wavelength interval& TOA flux & Observed variability from mean & Ref.\\
\multicolumn{1}{c}{(nm)} & (W~m$^{-2}$) & & \\
\hline
0.7--10 (X-rays)              &  0.131 & +100$\times$ (flares); $\pm$20\% (rotation); $\pm$20\% (cycle) & 1,2\\
10--40                        &  0.110 & +100$\times$ (flares); $\pm$20\% (rotation); $\pm$20\% (cycle) & 1,2\\
40--92                        &  0.033 & +10$\times$ (flares); no rotation \& cycle information & 3 \\
92--118                       &  0.019 & +30$\times$ (flares); no rotation \& cycle information & 4 \\
0.7--118 (XUV)                &  0.293 & +100$\times$ (flares); $\pm$20\% (rotation); $\pm$20\% (cycle) & 1,2\\
118--170 (FUV)                &  0.147 & +10$\times$ (flares); no rotation \& cycle information & 5 \\
H Ly$\alpha$ (122 nm)         &  0.130 & +10$\times$ (flares); no rotation \& cycle information & 5 \\
0.7--30000 ($S_{\rm Prox b}$) &  877   & +15\% (flares); $\pm$0.5\% (rotation); $\pm$0.5\% (cycle) & 6,2\\
\hline
\hline
\end{tabular}
\label{tab:XUV}

References: 1: \citet{Guedeletal2004}; 2: \citet{Wea17}; 3: \citet{Mea06}; 4: 
\citet{Christianetal2004}; 5: \citet{LoydFrance2014}, Loyd (priv. comm.); 6: \citet{Dea16}
\end{table*}

\section{IR excess} \label{IRexcess}

The flux residuals in Fig. \ref{fig:IRSEDfit} show a clear systematic offset beyond 
$\approx$2 $\mu$m, with the observed flux being $\sim$20\% larger than 
model predictions. This systematic difference can be interpreted as a mid-to-near IR excess 
associated with the Proxima system, which, to our knowledge, has not been pointed out before. 
A possible physical explanation is the presence of dust grains, in what could be a warm 
ring close to the star, scattering the light from Proxima. The presence of such a dust 
reservoir could be leftover from the formation process of the planetary system around 
Proxima. Worth noting here is the K0 planet-host HD 69830 \citep{Lea06}, which was found 
to have a mid-IR excess ($\sim$50\% over photosphere) and was interpreted by \citet{Bea11} 
as caused by small dust grains within 1 AU of the star. While no other warm disk around an 
old M dwarf has been reported, a cold resolved debris disk (an analog to the Kuiper Belt 
of our Solar System) was found by \citet{Lea12} with {\em Herschel} Space Observatory 
observations of GJ 581. Unfortunately, no far-IR measurements of Proxima are available 
to investigate the presence of a cold debris disk, which could lend additional support to 
the explanation of the mid-to-near IR excess that we find.

The systematic trend of the residuals could alternatively be related to certain shortcomings
of the theoretical models. However, this is rather unlikely, as a $\sim$20\% flux deficit 
is very significant and would have been identified before in other stars \citep{Mea15}. In 
addition, one could think that the differences are related to the SED fitting procedure. A 
higher $T_{\rm eff}$ value could yield fluxes in better agreement. Being in the Rayleigh-Jeans 
regime, this would mean a $\sim$20\% increase in temperature of $\sim$600 K. The other 
possibility is to assume a larger angular diameter by $\sim$10\%. Neither option 
can be valid because of the existence of strong constraints coming from the HST/STIS flux 
at 1~$\mu$m and from the interferometric angular diameter determination. Finally, 
one could also consider a heavily spotted stellar surface (i.e., a hotter photosphere and 
a significant fraction of cooler spot areal coverage) that could result in a SED with an 
apparent IR flux excess. However, a IR flux excess that becomes significant at 
$\approx$2 $\mu$m would require an unrealistically low spot temperature value 
(T$_{\rm spot}<1500$ K, $T_{\rm phot}-T_{\rm spot}>1500 K$; c.f., \citealt{B05}). Thus, 
given the lack of an alternative explanation consistent with the data and model fits, we 
find that the most likely cause of the IR excess is scattering of light from warm dust 
particles close to the star.

\section{Physical and radiative properties} \label{Lbol}

From the full spectral energy distribution of Proxima we can estimate its radiative
parameters. The integration of the total flux (from X-rays to 30~$\mu$m) yields a 
flux of $2.86\times10^{-8}$ erg s$^{-1}$ cm$^{-2}$. The uncertainty of this
value should be mostly driven by the uncertainty in the HST/STIS spectrophotometric 
measurements and of the IR fit. For the former, the absolute flux scale is found to be 
better than 5\%, and possibly better than 3\% \citep{Bea14}. For the IR, given the 
discussion on the quality of the fit, we also adopt an uncertainty of 5\%. Therefore, 
the bolometric flux of Proxima is found to be 
$F_{\rm bol} = (2.86\pm0.14)\times10^{-8}$ erg s$^{-1}$ cm$^{-2}$. This value and the 
angular diameter in Table \ref{tab:fitresults} lead to an effective temperature value of 
$T_{\rm eff} = 2980\pm80$~K. The difference from the value in Table \ref{tab:fitresults} 
arises because in this calculation we consider the full wavelength range, 
not only the interval beyond 1~$\mu$m. In other words, the optical flux of Proxima is 
lower than that of a 3000 K model and, hence, results in a lower $T_{\rm eff}$. 

Finally, we calculate the bolometric luminosity by using the stellar parallax of Proxima 
from \citet{Bea99} and a solar luminosity value from IAU 2015 Resolution B3 on 
Recommended Nominal Conversion Constants for Selected Solar and Planetary Properties 
(\url{https://www.iau.org/administration/resolutions/general_assemblies/}). The bolometric
luminosity of Proxima is, thus, $L_{\rm bol} = (5.80\pm0.30)\times10^{30}$ erg s$^{-1}$ or
$L_{\rm bol} = 0.00151\pm0.0008$ L$_{\odot}$.

\section{Time evolution of the flux received by Proxima b} \label{XUV}

\subsection{Bolometric flux}

\begin{figure}
\centering
\includegraphics[width=\columnwidth]{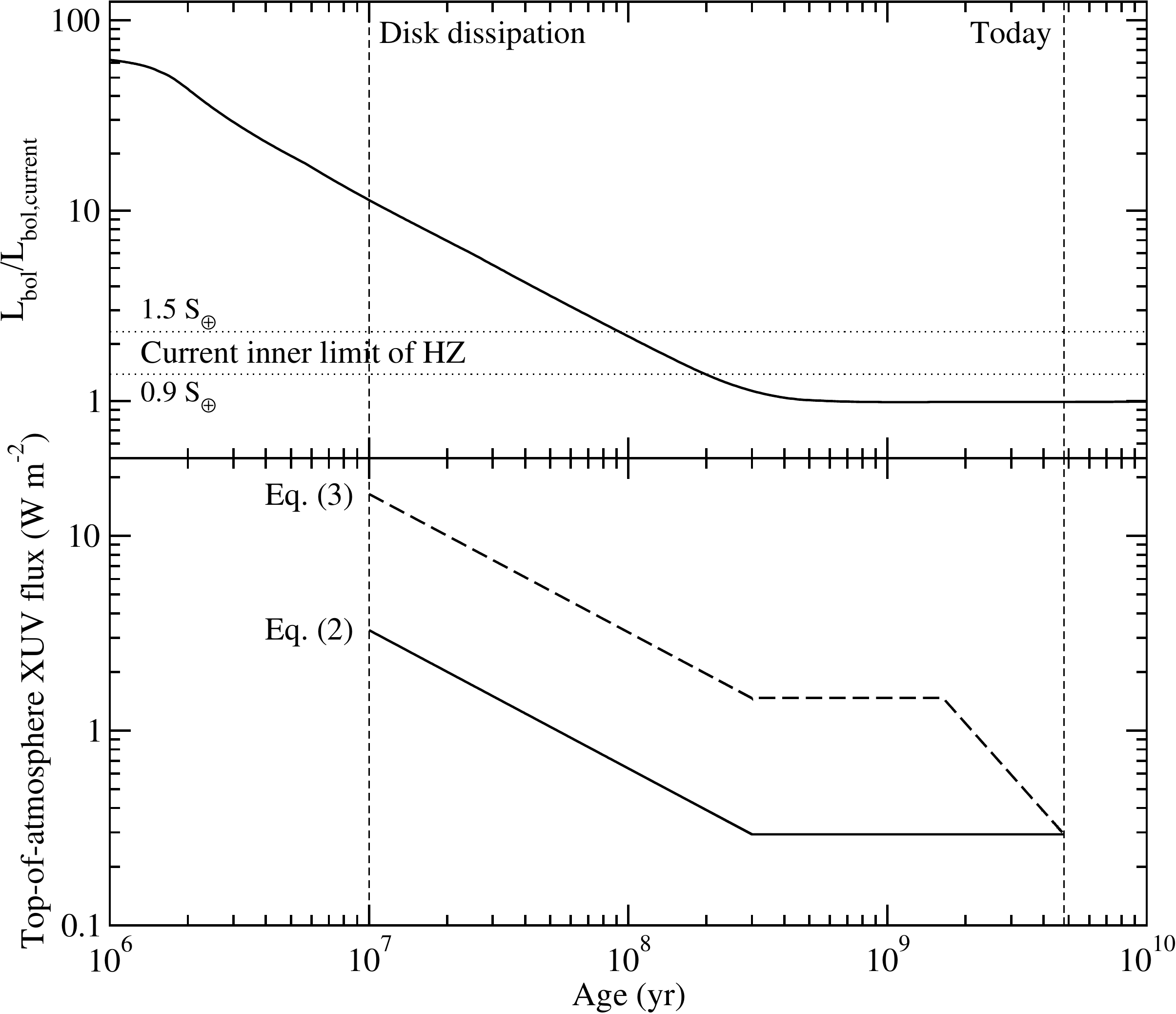}
\caption{{\em Top:} Evolution of the bolometric luminosity of a 0.120-M$_{\odot}$ star 
as predicted by the models of \citet{Baraffeetal15} and normalized to today's Proxima 
luminosity $L_{\rm bol} = 0.00151$~L$_{\odot}$. Marked with vertical dashed lines are 
today's Proxima age (4.8 Gyr) and the approximate age at which the Proxima protoplanetary 
disk may have dissipated \citep[10 Myr,][]{PM16}. The horizontal dotted lines mark the 
current inner limit of the habitable zone according to different assumptions on the spin 
rate of the planet \citep{K13}: synchronous, i.e., 1.5 times Earth's irradiance (S$_\oplus$); 
asynchronous, i.e., 0.9 S$_\oplus$. {\em Bottom:} XUV flux evolution calculated using the 
two prescriptions explained in Sects. \ref{sec:sat} (solid) and \ref{sec:pow} (thick dashed). 
Vertical dashed lines mark the same features as in the top panel.}
\label{fig:track}
\end{figure}

The total flux evolution of Proxima can be estimated from theoretical evolutionary
model calculations. We employed the recent models of \citet{Baraffeetal15} that are well
suited for very low mass stars and include the most up-to-date physical ingredients. 
We linearly interpolated the evolutionary tracks from the models corresponding to 0.110 
and 0.130~M$_{\odot}$ to find a good match of the model predictions with our determined 
values for $L_{\rm bol}$ and $T_{\rm eff}$ at the estimated age of the star of 
4.8 Gyr \citep{Bazotetal2016}. A stellar mass of 0.120~M$_{\odot}$ yields the best 
simultaneous agreement of all parameters within the corresponding uncertainties, resulting
in values of $L_{\rm bol} = 0.00150$~L$_{\odot}$ and $T_{\rm eff} = 2980$~K. A formal
uncertainty of 0.003~M$_{\odot}$ can be estimated from the errors associated to 
$L_{\rm bol}$ and $T_{\rm eff}$. This is obviously a model-dependent estimate and 
no error in metallicity and $\log g$ was assumed. The evolutionary track of Proxima is shown 
in Fig. \ref{fig:track}, in normalized units of today's bolometric luminosity. At 10 Myr, 
the time when the protoplanetary disk may have dissipated \citep{PM16} and Proxima b 
became vulnerable to XUV radiation, the stellar luminosity was a factor of 10 larger than 
today. Thus, Proxima b spent some 90--200 Myr in an orbit interior to the stellar 
habitable zone and possibly in a runaway greenhouse state. A detailed discussion is 
provided by \citet{Rea16}.

\subsection{High-energy flux}

The evolution of the XUV flux of Proxima with time was addressed by \citet{Rea16}. Here 
we revisit the calculations by considering also the early evolution of $L_{\rm bol}$ as the 
star was contracting towards the Main Sequence. The XUV evolution of young M dwarfs is poorly
constrained but some tantalizing evidence exists indicating that the saturation limit of 
$\log (L_{\rm X}/L_{\rm bol}) \approx -4$ also applies to the pre-Main Sequence 
\citep{PF05}. As we show above, the bolometric luminosity should have experienced significant 
changes over the first few hundred Myr in the history of Proxima and therefore this needs 
to be properly taken into account in the calculations \citep[e.g.,][]{LB15}. Different XUV 
evolution laws are discussed in \citet{Rea16}. One considers a saturated emission state 
up to a certain age followed by a power law decrease to today's XUV flux. The other one 
considers that Proxima has shown saturated behaviour since its birth and until today. 
Observational evidence is still inconclusive as to which of these two XUV 
evolution scenarios is correct, and we hereby further consider them both. They should be
representative of the extreme cases bracketing the real evolution of Proxima over its
lifetime.

\subsubsection{Proxima is just at the end of the saturation phase} \label{sec:sat}

Proxima's current relative X-ray value is $\log (L_{\rm X}/L_{\rm bol}) = -3.83$, 
which is very similar to the average of the distribution for stars between 0.1 and 
0.2~M$_{\odot}$ and ages of 0.1 to 10 Myr found by \citet{PF05}. This circumstance 
suggests that Proxima may still be today in the saturated regime and 
that $\log (L_{\rm X}/L_{\rm bol}) = -3.83$ has been satisfied during its 
entire lifetime. This is in good agreement with the estimates of the saturation limit 
as determined from the equations in \citet{Reinersetal2014}, which extends up to a 
rotation period of $P_{\rm rot}\approx80$~d for a 0.146-R$_{\odot}$ star, very close to 
Proxima's current rotation period of $P_{\rm rot}=83$~d \citep{SMea16}.

To parameterize the bolometric flux, we consider the evolutionary model track and two 
different regimes, from 10 to 300 Myr and from 300 Myr to today. The evolution of the
stellar bolometric flux as a function of the age ($\tau$) in Myr can be approximated as
(see top panel of Fig. \ref{fig:track}):
\begin{equation}
L_{\rm bol}/L_{\rm bol, current}=57.38 \: \tau^{-0.71} \hspace*{3mm} 
\mbox{for 10 Myr $< \tau <$ 300 Myr}\nonumber
\end{equation}
\begin{equation}
L_{\rm bol}/L_{\rm bol, current}=1.000 \hspace*{11mm} \mbox{for 300 Myr $< \tau <$ 4800 Myr}
\end{equation}
Then, we assume that the XUV flux scales in the same way as the X-rays. This is an 
approximation because the hardness ratio of the XUV spectrum may have softened as the
star spun down. But in the absence of a better model, we used the evolution law from
X-rays as valid for the full XUV range, and, therefore, that $\log (L_{\rm XUV}/L_{\rm bol}) 
= -3.48$ has remained constant for the entire lifetime of Proxima. With this, and the 
expressions in Eq. (1), we find the following relationship for the top-of-atmosphere flux 
of Proxima b as a function of age (the current value is taken from Table \ref{tab:XUV}):
\begin{equation}
F_{\rm XUV}=16.81 \: \tau^{-0.71} \:\: \mbox{W~m$^{-2}$} \hspace*{3mm} 
\mbox{for 10 Myr $< \tau <$ 300 Myr}\nonumber
\end{equation}
\begin{equation}
F_{\rm XUV}=0.293 \:\:\mbox{W m$^{-2}$} \hspace*{13mm} \mbox{for 300 Myr $< \tau <$ 4800 Myr}
\end{equation}
The proposed evolution of the top-of-atmosphere flux received by Proxima b corresponding
to this scenario is illustrated in the bottom panel of Fig. \ref{fig:track} with a solid line.

\subsubsection{Proxima has evolved off saturation and is in the power law regime} \label{sec:pow}

As an alternative to the XUV evolution scenario above, one can consider the results 
of \citet{WD16}. Although based on 4 stars (among which is Proxima), the authors suggest that 
the X-ray evolution of fully convective stars is analogous to that of more massive Sun-like 
stars. In this case, to model the stage after saturation, we can adopt the relationship in 
\citet{Wea11} by which $L_{\rm X}/L_{\rm bol} \propto R_{\circ}^{-2.70} \propto P_{\rm 
rot}^{-2.70}$, where $R_{\circ} \equiv P_{\rm rot}/\tau_{\rm c}$ is the so-called Rossby 
number \citep{Nea84} and we assume the convective turnover time ($\tau_{\rm c}$) to be 
constant during the main sequence lifetime of Proxima. From this, we can further adopt 
\citet{MH08}, who find $P_{\rm rot} \propto \tau^{0.566}$, where $\tau$ is the stellar age, 
to obtain $L_{\rm X}/L_{\rm bol} \propto \tau^{-1.5}$. Thus, considering that 
$\log (L_{\rm X}/L_{\rm bol}) = -3.83$ at an age of 4.8~Gyr and that saturation of Sun-like 
stars occurs at an average value of $\log (L_{\rm X}/L_{\rm bol}) = -3.13$ \citep{Wea11}, we 
find that the end of saturation should have happened at an age of 1.64 Gyr.

As before, we further make the assumption that the total XUV flux follows the same 
evolution as the X-ray flux and we can write the relationship ($\tau$ in Myr):
\begin{equation}
F_{\rm XUV}=84.1 \: \tau^{-0.71} \:\: \mbox{W~m$^{-2}$} \hspace*{9mm} 
\mbox{for 10 Myr $< \tau <$ 300 Myr}\nonumber
\end{equation}
\begin{equation}
F_{\rm XUV}=1.47 \:\:\mbox{W m$^{-2}$} \hspace*{17.5mm} \mbox{for 300 Myr $< \tau <$ 1640 Myr} 
\nonumber
\end{equation}
\begin{equation}
F_{\rm XUV}=9.74\times10^4 \: \tau^{-1.5} \:\: \mbox{W~m$^{-2}$} \hspace*{2mm} 
\mbox{for 1640 Myr $< \tau <$ 4800 Myr}
\end{equation}
This proposed evolution is illustrated in the bottom panel of Fig. \ref{fig:track} with 
a thick dashed line.

\subsection{XUV dose and water loss estimates}

The integration of the XUV relationships presented here and the comparison with 
the equivalent relationship for the Sun and the Earth \citep[see][]{Rea16} indicates 
that the total XUV dose that Proxima b has received over its lifetime is between 8 and 
25 times greater than Earth's. But the most critical part may be the phase at 
which the atmosphere of Proxima b was in runaway greenhouse effect, in an orbit interior 
to the habitable zone. The amount of XUV irradiation during this period of time from about 
10 Myr until about 90--200 Myr could have caused an intense loss of water. To estimate the 
water loss, we proceeded as in \citet{Rea16} and \citet{Bolmont2017}. We use the same 
units for the water loss as in those articles: 1~$EO_H$ corresponds to the Earth ocean's 
worth of hydrogen. We also took into account the revised smaller mass for Proxima, but 
this has no significant impact on the calculations. With our model, we can estimate 
the current volatile losses of Proxima b: the hydrogen loss is of 0.003~$EO_H$/Myr, which 
corresponds to $1.5\times10^7$~g~s$^{-1}$, the oxygen loss is 0.009~$EO_H$/Myr, which 
corresponds to $4.3\times10^7$~g~s$^{-1}$.

Table~\ref{tab:Hloss} summarizes the results for the two prescriptions given in 
Eqs. (2) and (3). The parameterization of the XUV flux evolution given by Eq. (2) differs from 
the one used in \citet{Rea16} as follows: it is higher during the first 100~Myr but lower 
by a factor $\sim$2.7 during the following few Gyr. This has two consequences on the water 
loss: 1) during the runaway phase, and more especially during the first 100~Myr, the loss is 
more intense than in \citet{Rea16}, and 2), on the long term, the total loss is lower. 
The parameterization of the evolution of the XUV flux given by Eq.~(3) leads to higher XUV 
fluxes throughout the entire lifetime of Proxima b when compared with \citet{Rea16}.

\begin{table}
\centering
\caption{Hydrogen loss from Proxima b for different XUV prescriptions. The 
following assumptions are made: the initial time is 10~Myr, the mass and radius of Proxima b 
are 1.3~M$_\oplus$ and 1.07~$R_\oplus$, respectively, and the initial water content is 
considered infinite.}
\begin{tabular}{cccc}
\hline
\hline
XUV prescription & \multicolumn{3}{c}{H loss ($EO_H$)}\\
 & HZ (1.5~S$_\oplus$)   & HZ (0.9~S$_\oplus$) & Lifetime \\
 & 90~Myr           & 200~Myr       & 4.8~Gyr \\
\hline
Eq. (2) & 0.47  & 0.90  & 15.6  \\
Eq. (3) & 1.07  & 1.98  & 24.4  \\
\hline
\hline
\end{tabular}
\label{tab:Hloss}
\end{table}

If we assume synchronous rotation, our estimates indicate that Proxima b could have lost 
from 0.47~$EO_H$ (Eq.~2) to 1.07~$EO_H$ (Eq.~3) between 10~Myr and 90~Myr, when it 
reached the inner edge of the habitable zone at 1.5~S$_{\oplus}$ \citep{Kopparapu2016}. Our 
new calculations therefore suggest that, during that time, Proxima b may have lost more 
water than previously estimated by \citet{Rea16}, by about a factor 1.25 to 3. 
Assuming non-synchronous rotation, the amount of water lost could range from 
0.9~$EO_H$ (Eq.~2) to 1.91~$EO_H$ (Eq.~3) between 10~Myr and 200~Myr, when it reached 
the habitable zone inner edge at 0.9~S$_{\oplus}$ \citep{Kopparapu2014}. The 
estimate obtained with the prescription of Eq.~(2) is about the same value as previously 
provided by \citet{Rea16} while the calculation with Eq. (3) is about a factor of 2 
larger.

In spite of the strong volatile losses ($\sim$0.5--2~$EO_H$), the planet could still have a 
significant amount of water reservoir when it entered the habitable zone depending on the 
initial content. What could have occurred beyond this point is quite uncertain. If we assume 
that the water loss processes were still active upon entering the habitable zone, we 
find that Proxima b could have lost up to 15--25~$EO_H$ during its lifetime. However, this 
needs to be considered an extreme upper limit because the volatile loss mechanisms would 
probably be significantly less efficient under such conditions \citep[see discussion 
in][]{Rea16}. 

\section{Conclusions} \label{concl}

This paper presents a full analysis of the SED of Proxima, covering X-rays to 
mid-IR, with the goal of providing useful input to study the atmosphere of Proxima b. 
We made use of measurements covering different wavelength intervals and acquired with 
various facilities (see Tables \ref{tab:facilities} and \ref{tab:IRflux}) to determine 
top-of-atmosphere fluxes from 0.7 to 30000 nm, in steps of widths ranging from 0.05 to 
10 nm depending on the wavelength range. Where spectrophotometric measurements were 
unavailable, we made use of theoretical models fitted using all available 
constraints. With the full spectral energy distribution and the available trigonometric 
distance, we could calculate the bolometric luminosity and the effective temperature. 
Also, Proxima has a quite accurate interferometric angular diameter measurement and this 
was used both to constrain the SED fit in the IR and to provide an empirical determination 
of the stellar radius. Interestingly, the fit of the IR SED revealed a flux 
excess $\approx$20\% from Proxima.  While the origin of this excess is uncertain, the most 
natural explanation is light scattering by dust particles in the Proxima system; 
additional observations can better ascertain the nature of the excess. The stellar mass 
was estimated by comparison with evolutionary models using the constraints provided by 
the radiative properties of Proxima. All the resulting fundamental parameters are 
summarized in Table \ref{tab:summary}. 

\begin{table}
\centering
\caption{Summary of fundamental properties of Proxima.}
\begin{tabular}{lc}
\hline
\hline
Parameter & Value \\
\hline
$M$ (M$_{\odot}$)                      & $0.120\pm0.003\tablefootmark{a}$ \\
$R$ (R$_{\odot}$)                      & $0.146\pm0.007$ \\
$T_{\rm eff}$ (K)                      & $2980\pm80$  \\
$F_{\rm bol}$ (erg s$^{-1}$ cm$^{-2}$) & $2.86\pm0.14\times10^{-8}$ \\
$L_{\rm bol}$ (L$_{{\rm bol}\odot}$)   & $0.00151\pm0.00008$  \\
$<\log (L_{\rm X}/L_{\rm bol})>$       & $-3.83$  \\
$<\log (L_{\rm XUV}/L_{\rm bol})>$     & $-3.48$  \\
Age (Gyr)\tablefootmark{b}             & 4.8 \\
\hline
\hline
\end{tabular}
\tablefoot{
\tablefoottext{a}{Model dependent, from errors in $T_{\rm eff}$ and $L_{\rm bol}$ but assuming
no error in age and metallicity.}
\tablefoottext{b}{From \citet{Bazotetal2016}.}}
\label{tab:summary}
\end{table}

Proxima is a benchmark star, not only for us to understand the stellar lower main sequence, 
but also, since the discovery of Proxima b, to study its habitable planet candidate.
As discussed by \citet{Rea16}, to determine the habitability of the planet it is 
essential to analyze the volatile loss processes that may affect its atmosphere, both
currently and in the past. The detailed spectral energy distribution for Proxima presented 
here and the newly proposed XUV flux time evolution laws should help to provide the 
necessary constraints to model and interpret future observations of the nearest potentially 
habitable planet outside the Solar System.

\begin{acknowledgements}
We are grateful to Rodrigo Luger for pointing out the increased XUV flux in the early
evolution of Proxima, and to Parke Loyd for assistance with the flare characterization
in the FUV range. We also gratefully acknowledge the insightful comments and 
suggestions by an anonymous referee. I.~R. acknowledges support by the Spanish Ministry 
of Economy and Competitiveness (MINECO) and the Fondo Europeo de Desarrollo Regional 
(FEDER) through grant ESP2016-80435-C2-1-R, as well as the support of the Generalitat 
de Catalunya/CERCA programme. M.~D.~G. and T.~S.~B. acknowledge generous support provided 
by NASA through grant number GO-13776 from the Space Telescope Science Institute, which 
is operated by AURA, Inc., under NASA contract NAS5-26555. E.~B. acknowledges funding by 
the European Research Council through ERC grant SPIRE 647383.
\end{acknowledgements}

   \bibliographystyle{aa} 
   \bibliography{biblio} 

\begin{thebibliography}{75}
\expandafter\ifx\csname natexlab\endcsname\relax\def\natexlab#1{#1}\fi

\bibitem[{{Anglada-Escud{\'e}} {et~al.}(2016){Anglada-Escud{\'e}}, {Amado},
  {Barnes}, {Berdi{\~n}as}, {Butler}, {Coleman}, {de La Cueva}, {Dreizler},
  {Endl}, {Giesers}, {Jeffers}, {Jenkins}, {Jones}, {Kiraga}, {K{\"u}rster},
  {L{\'o}pez-Gonz{\'a}lez}, {Marvin}, {Morales}, {Morin}, {Nelson}, {Ortiz},
  {Ofir}, {Paardekooper}, {Reiners}, {Rodr{\'{\i}}guez},
  {Rodr{\'{\i}}guez-L{\'o}pez}, {Sarmiento}, {Strachan}, {Tsapras}, {Tuomi}, \&
  {Zechmeister}}]{AngladaEscudeetal2016}
{Anglada-Escud{\'e}}, G., {Amado}, P.~J., {Barnes}, J., {et~al.} 2016, \nat,
  536, 437

\bibitem[{{Aumann} \& {Probst}(1991)}]{AP91}
{Aumann}, H.~H. \& {Probst}, R.~G. 1991, \apj, 368, 264

\bibitem[{{Ayres}(2010)}]{Ayres2010}
{Ayres}, T.~R. 2010, \apjs, 187, 149

\bibitem[{{Baraffe} {et~al.}(2015){Baraffe}, {Homeier}, {Allard}, \&
  {Chabrier}}]{Baraffeetal15}
{Baraffe}, I., {Homeier}, D., {Allard}, F., \& {Chabrier}, G. 2015, \aap, 577,
  A42

\bibitem[{{Barnes} {et~al.}(2016){Barnes}, {Deitrick}, {Luger}, {Driscoll},
  {Quinn}, {Fleming}, {Guyer}, {McDonald}, {Meadows}, {Arney}, {Crisp},
  {Domagal-Goldman}, {Lincowski}, {Lustig-Yaeger}, \& {Schwieterman}}]{Bea16}
{Barnes}, R., {Deitrick}, R., {Luger}, R., {et~al.} 2016, ArXiv e-prints

\bibitem[{{Bazot} {et~al.}(2016){Bazot}, {Christensen-Dalsgaard}, {Gizon}, \&
  {Benomar}}]{Bazotetal2016}
{Bazot}, M., {Christensen-Dalsgaard}, J., {Gizon}, L., \& {Benomar}, O. 2016,
  \mnras, 460, 1254

\bibitem[{{Beichman} {et~al.}(2011){Beichman}, {Lisse}, {Tanner}, {Bryden},
  {Akeson}, {Ciardi}, {Boden}, {Dodson-Robinson}, {Salyk}, \& {Wyatt}}]{Bea11}
{Beichman}, C.~A., {Lisse}, C.~M., {Tanner}, A.~M., {et~al.} 2011, \apj, 743,
  85

\bibitem[{{Benedict} {et~al.}(1999){Benedict}, {McArthur}, {Chappell}, {Nelan},
  {Jefferys}, {van Altena}, {Lee}, {Cornell}, {Shelus}, {Hemenway}, {Franz},
  {Wasserman}, {Duncombe}, {Story}, {Whipple}, \& {Fredrick}}]{Bea99}
{Benedict}, G.~F., {McArthur}, B., {Chappell}, D.~W., {et~al.} 1999, \aj, 118,
  1086

\bibitem[{{Berdyugina}(2005)}]{B05}
{Berdyugina}, S.~V. 2005, Living Reviews in Solar Physics, 2, 8

\bibitem[{{Bessell} \& {Murphy}(2012)}]{BM12}
{Bessell}, M. \& {Murphy}, S. 2012, \pasp, 124, 140

\bibitem[{{Bessell}(1991)}]{B91}
{Bessell}, M.~S. 1991, \aj, 101, 662

\bibitem[{{Bessell} {et~al.}(1998){Bessell}, {Castelli}, \& {Plez}}]{Bea98}
{Bessell}, M.~S., {Castelli}, F., \& {Plez}, B. 1998, \aap, 333, 231

\bibitem[{{Bohlin} {et~al.}(2014){Bohlin}, {Gordon}, \& {Tremblay}}]{Bea14}
{Bohlin}, R.~C., {Gordon}, K.~D., \& {Tremblay}, P.-E. 2014, \pasp, 126, 711

\bibitem[{{Bolmont} {et~al.}(2017){Bolmont}, {Selsis}, {Owen}, {Ribas},
  {Raymond}, {Leconte}, \& {Gillon}}]{Bolmont2017}
{Bolmont}, E., {Selsis}, F., {Owen}, J.~E., {et~al.} 2017, \mnras, 464, 3728

\bibitem[{{Christian} {et~al.}(2004){Christian}, {Mathioudakis}, {Bloomfield},
  {Dupuis}, \& {Keenan}}]{Christianetal2004}
{Christian}, D.~J., {Mathioudakis}, M., {Bloomfield}, D.~S., {Dupuis}, J., \&
  {Keenan}, F.~P. 2004, \apj, 612, 1140

\bibitem[{{Cohen} {et~al.}(2003){Cohen}, {Wheaton}, \& {Megeath}}]{Coea03}
{Cohen}, M., {Wheaton}, W.~A., \& {Megeath}, S.~T. 2003, \aj, 126, 1090

\bibitem[{{Cutri} {et~al.}(2003){Cutri}, {Skrutskie}, {van Dyk}, {Beichman},
  {Carpenter}, {Chester}, {Cambresy}, {Evans}, {Fowler}, {Gizis}, {Howard},
  {Huchra}, {Jarrett}, {Kopan}, {Kirkpatrick}, {Light}, {Marsh}, {McCallon},
  {Schneider}, {Stiening}, {Sykes}, {Weinberg}, {Wheaton}, {Wheelock}, \&
  {Zacarias}}]{Cea03}
{Cutri}, R.~M., {Skrutskie}, M.~F., {van Dyk}, S., {et~al.} 2003, VizieR Online
  Data Catalog, 2246

\bibitem[{{Davenport} {et~al.}(2016){Davenport}, {Kipping}, {Sasselov},
  {Matthews}, \& {Cameron}}]{Dea16}
{Davenport}, J.~R.~A., {Kipping}, D.~M., {Sasselov}, D., {Matthews}, J.~M., \&
  {Cameron}, C. 2016, \apjl, 829, L31

\bibitem[{{Demory} {et~al.}(2009){Demory}, {S{\'e}gransan}, {Forveille},
  {Queloz}, {Beuzit}, {Delfosse}, {di Folco}, {Kervella}, {Le Bouquin},
  {Perrier}, {Benisty}, {Duvert}, {Hofmann}, {Lopez}, \& {Petrov}}]{Dea09}
{Demory}, B.-O., {S{\'e}gransan}, D., {Forveille}, T., {et~al.} 2009, \aap,
  505, 205

\bibitem[{{Egan} {et~al.}(2003){Egan}, {Price}, {Kraemer}, {Mizuno}, {Carey},
  {Wright}, {Engelke}, {Cohen}, \& {Gugliotti}}]{Eea03}
{Egan}, M.~P., {Price}, S.~D., {Kraemer}, K.~E., {et~al.} 2003, VizieR Online
  Data Catalog, 5114

\bibitem[{{Egan} {et~al.}(1999){Egan}, {Price}, {Moshir}, {Cohen}, \&
  {Tedesco}}]{Eea99}
{Egan}, M.~P., {Price}, S.~D., {Moshir}, M.~M., {Cohen}, M., \& {Tedesco}, E.
  1999, NASA STI/Recon Technical Report N

\bibitem[{{Frogel} {et~al.}(1972){Frogel}, {Kleinmann}, {Kunkel}, {Ney}, \&
  {Strecker}}]{Fea72}
{Frogel}, J.~A., {Kleinmann}, D.~E., {Kunkel}, W., {Ney}, E.~P., \& {Strecker},
  D.~W. 1972, \pasp, 84, 581

\bibitem[{{Fr{\"o}hlich}(2012)}]{F12}
{Fr{\"o}hlich}, C. 2012, Surveys in Geophysics, 33, 453

\bibitem[{{Fuhrmeister} {et~al.}(2011){Fuhrmeister}, {Lalitha}, {Poppenhaeger},
  {Rudolf}, {Liefke}, {Reiners}, {Schmitt}, \& {Ness}}]{Fuhrmeisteretal2011}
{Fuhrmeister}, B., {Lalitha}, S., {Poppenhaeger}, K., {et~al.} 2011, \aap, 534,
  A133

\bibitem[{{Gautier} {et~al.}(2007){Gautier}, {Rieke}, {Stansberry}, {Bryden},
  {Stapelfeldt}, {Werner}, {Beichman}, {Chen}, {Su}, {Trilling}, {Patten}, \&
  {Roellig}}]{Gea07}
{Gautier}, III, T.~N., {Rieke}, G.~H., {Stansberry}, J., {et~al.} 2007, \apj,
  667, 527

\bibitem[{{Gezari} {et~al.}(1999){Gezari}, {Pitts}, \& {Schmitz}}]{Gea99}
{Gezari}, D.~Y., {Pitts}, P.~S., \& {Schmitz}, M. 1999, VizieR Online Data
  Catalog, 2225

\bibitem[{{Gliese} \& {Jahreiss}(2015)}]{GJ15}
{Gliese}, W. \& {Jahreiss}, H. 2015, VizieR Online Data Catalog, 5035

\bibitem[{{G{\"u}del} {et~al.}(2004){G{\"u}del}, {Audard}, {Reale}, {Skinner},
  \& {Linsky}}]{Guedeletal2004}
{G{\"u}del}, M., {Audard}, M., {Reale}, F., {Skinner}, S.~L., \& {Linsky},
  J.~L. 2004, \aap, 416, 713

\bibitem[{{Guinan} {et~al.}(2003){Guinan}, {Ribas}, \&
  {Harper}}]{Guinanetal2003}
{Guinan}, E.~F., {Ribas}, I., \& {Harper}, G.~M. 2003, \apj, 594, 561

\bibitem[{{Helou} \& {Walker}(1988)}]{IRAS}
{Helou}, G. \& {Walker}, D.~W., eds. 1988, {Infrared astronomical satellite
  (IRAS) catalogs and atlases. Volume 7: The small scale structure catalog},
  Vol.~7, 1--265

\bibitem[{{Herrero} {et~al.}(2016){Herrero}, {Ribas}, {Jordi}, {Morales},
  {Perger}, \& {Rosich}}]{Hea16}
{Herrero}, E., {Ribas}, I., {Jordi}, C., {et~al.} 2016, \aap, 586, A131

\bibitem[{{Hudson}(1971)}]{H71}
{Hudson}, R.~D. 1971, Reviews of Geophysics and Space Physics, 9, 305

\bibitem[{{Jao} {et~al.}(2014){Jao}, {Henry}, {Subasavage}, {Winters}, {Gies},
  {Riedel}, \& {Ianna}}]{Jea14}
{Jao}, W.-C., {Henry}, T.~J., {Subasavage}, J.~P., {et~al.} 2014, \aj, 147, 21

\bibitem[{{Jarrett} {et~al.}(2011){Jarrett}, {Cohen}, {Masci}, {Wright},
  {Stern}, {Benford}, {Blain}, {Carey}, {Cutri}, {Eisenhardt}, {Lonsdale},
  {Mainzer}, {Marsh}, {Padgett}, {Petty}, {Ressler}, {Skrutskie}, {Stanford},
  {Surace}, {Tsai}, {Wheelock}, \& {Yan}}]{Jea11}
{Jarrett}, T.~H., {Cohen}, M., {Masci}, F., {et~al.} 2011, \apj, 735, 112

\bibitem[{{Joint IRAS Science}(1994)}]{IRAS94}
{Joint IRAS Science}, W.~G. 1994, VizieR Online Data Catalog, 2125

\bibitem[{{Kopp} \& {Lean}(2011)}]{KL11}
{Kopp}, G. \& {Lean}, J.~L. 2011, \grl, 38, L01706

\bibitem[{{Kopparapu}(2013)}]{K13}
{Kopparapu}, R.~K. 2013, \apjl, 767, L8

\bibitem[{{Kopparapu} {et~al.}(2014){Kopparapu}, {Ramirez}, {SchottelKotte},
  {Kasting}, {Domagal-Goldman}, \& {Eymet}}]{Kopparapu2014}
{Kopparapu}, R.~K., {Ramirez}, R.~M., {SchottelKotte}, J., {et~al.} 2014,
  \apjl, 787, L29

\bibitem[{{Kopparapu} {et~al.}(2016){Kopparapu}, {Wolf}, {Haqq-Misra}, {Yang},
  {Kasting}, {Meadows}, {Terrien}, \& {Mahadevan}}]{Kopparapu2016}
{Kopparapu}, R.~k., {Wolf}, E.~T., {Haqq-Misra}, J., {et~al.} 2016, \apj, 819,
  84

\bibitem[{{Lestrade} {et~al.}(2012){Lestrade}, {Matthews}, {Sibthorpe},
  {Kennedy}, {Wyatt}, {Bryden}, {Greaves}, {Thilliez}, {Moro-Mart{\'{\i}}n},
  {Booth}, {Dent}, {Duch{\^e}ne}, {Harvey}, {Horner}, {Kalas}, {Kavelaars},
  {Phillips}, {Rodriguez}, {Su}, \& {Wilner}}]{Lea12}
{Lestrade}, J.-F., {Matthews}, B.~C., {Sibthorpe}, B., {et~al.} 2012, \aap,
  548, A86

\bibitem[{{Lindler} \& {Heap}(2008)}]{LH08}
{Lindler}, D. \& {Heap}, S.~R. 2008,
  \url{https://archive.stsci.edu/pub/hlsp/stisngsl/aaareadme.pdf}

\bibitem[{{Linsky} {et~al.}(2014){Linsky}, {Fontenla}, \&
  {France}}]{Linskyetal2014}
{Linsky}, J.~L., {Fontenla}, J., \& {France}, K. 2014, \apj, 780, 61

\bibitem[{{Lovis} {et~al.}(2006){Lovis}, {Mayor}, {Pepe}, {Alibert}, {Benz},
  {Bouchy}, {Correia}, {Laskar}, {Mordasini}, {Queloz}, {Santos}, {Udry},
  {Bertaux}, \& {Sivan}}]{Lea06}
{Lovis}, C., {Mayor}, M., {Pepe}, F., {et~al.} 2006, \nat, 441, 305

\bibitem[{{Loyd} \& {France}(2014)}]{LoydFrance2014}
{Loyd}, R.~O.~P. \& {France}, K. 2014, \apjs, 211, 9

\bibitem[{{Luger} \& {Barnes}(2015)}]{LB15}
{Luger}, R. \& {Barnes}, R. 2015, Astrobiology, 15, 119

\bibitem[{{Mamajek} \& {Hillenbrand}(2008)}]{MH08}
{Mamajek}, E.~E. \& {Hillenbrand}, L.~A. 2008, \apj, 687, 1264

\bibitem[{{Mann} {et~al.}(2015){Mann}, {Feiden}, {Gaidos}, {Boyajian}, \& {von
  Braun}}]{Mea15}
{Mann}, A.~W., {Feiden}, G.~A., {Gaidos}, E., {Boyajian}, T., \& {von Braun},
  K. 2015, \apj, 804, 64

\bibitem[{{Mann} \& {von Braun}(2015)}]{Mv15}
{Mann}, A.~W. \& {von Braun}, K. 2015, \pasp, 127, 102

\bibitem[{{Marty}(2012)}]{M12}
{Marty}, B. 2012, Earth and Planetary Science Letters, 313, 56

\bibitem[{{Mould} \& {Hyland}(1976)}]{MH76}
{Mould}, J.~R. \& {Hyland}, A.~R. 1976, \apj, 208, 399

\bibitem[{{Mullan} {et~al.}(2006){Mullan}, {Mathioudakis}, {Bloomfield}, \&
  {Christian}}]{Mea06}
{Mullan}, D.~J., {Mathioudakis}, M., {Bloomfield}, D.~S., \& {Christian}, D.~J.
  2006, \apjs, 164, 173

\bibitem[{{Noyes} {et~al.}(1984){Noyes}, {Hartmann}, {Baliunas}, {Duncan}, \&
  {Vaughan}}]{Nea84}
{Noyes}, R.~W., {Hartmann}, L.~W., {Baliunas}, S.~L., {Duncan}, D.~K., \&
  {Vaughan}, A.~H. 1984, \apj, 279, 763

\bibitem[{{Passegger} {et~al.}(2016){Passegger}, {Wende-von Berg}, \&
  {Reiners}}]{Pea16}
{Passegger}, V.~M., {Wende-von Berg}, S., \& {Reiners}, A. 2016, \aap, 587, A19

\bibitem[{{Pecaut} \& {Mamajek}(2016)}]{PM16}
{Pecaut}, M.~J. \& {Mamajek}, E.~E. 2016, \mnras, 461, 794

\bibitem[{{Plucinsky} {et~al.}(2017){Plucinsky}, {Beardmore}, {Foster},
  {Haberl}, {Miller}, {Pollock}, \& {Sembay}}]{Pea17}
{Plucinsky}, P.~P., {Beardmore}, A.~P., {Foster}, A., {et~al.} 2017, \aap, 597,
  A35

\bibitem[{{Preibisch} \& {Feigelson}(2005)}]{PF05}
{Preibisch}, T. \& {Feigelson}, E.~D. 2005, \apjs, 160, 390

\bibitem[{{Press} {et~al.}(1992){Press}, {Teukolsky}, {Vetterling}, \&
  {Flannery}}]{NR92}
{Press}, W.~H., {Teukolsky}, S.~A., {Vetterling}, W.~T., \& {Flannery}, B.~P.
  1992, {Numerical recipes in FORTRAN. The art of scientific computing}

\bibitem[{{Ranjan} \& {Sasselov}(2016)}]{RS16}
{Ranjan}, S. \& {Sasselov}, D.~D. 2016, Astrobiology, 16, 68

\bibitem[{{Redfield} {et~al.}(2002){Redfield}, {Linsky}, {Ake}, {Ayres},
  {Dupree}, {Robinson}, {Wood}, \& {Young}}]{Redfieldetal2002}
{Redfield}, S., {Linsky}, J.~L., {Ake}, T.~B., {et~al.} 2002, \apj, 581, 626

\bibitem[{{Reid}(1982)}]{R82}
{Reid}, N. 1982, \mnras, 201, 51

\bibitem[{{Reiners} {et~al.}(2014){Reiners}, {Sch{\"u}ssler}, \&
  {Passegger}}]{Reinersetal2014}
{Reiners}, A., {Sch{\"u}ssler}, M., \& {Passegger}, V.~M. 2014, \apj, 794, 144

\bibitem[{{Ribas} {et~al.}(2016){Ribas}, {Bolmont}, {Selsis}, {Reiners},
  {Leconte}, {Raymond}, {Engle}, {Guinan}, {Morin}, {Turbet}, {Forget}, \&
  {Anglada-Escud{\'e}}}]{Rea16}
{Ribas}, I., {Bolmont}, E., {Selsis}, F., {et~al.} 2016, \aap, 596, A111

\bibitem[{{Ribas} {et~al.}(2005){Ribas}, {Guinan}, {G{\"u}del}, \&
  {Audard}}]{Ribasetal2005}
{Ribas}, I., {Guinan}, E.~F., {G{\"u}del}, M., \& {Audard}, M. 2005, \apj, 622,
  680

\bibitem[{{Rieke} {et~al.}(2004){Rieke}, {Young}, {Engelbracht}, {Kelly},
  {Low}, {Haller}, {Beeman}, {Gordon}, {Stansberry}, {Misselt}, {Cadien},
  {Morrison}, {Rivlis}, {Latter}, {Noriega-Crespo}, {Padgett}, {Stapelfeldt},
  {Hines}, {Egami}, {Muzerolle}, {Alonso-Herrero}, {Blaylock}, {Dole}, {Hinz},
  {Le Floc'h}, {Papovich}, {P{\'e}rez-Gonz{\'a}lez}, {Smith}, {Su}, {Bennett},
  {Frayer}, {Henderson}, {Lu}, {Masci}, {Pesenson}, {Rebull}, {Rho}, {Keene},
  {Stolovy}, {Wachter}, {Wheaton}, {Werner}, \& {Richards}}]{MIPS04}
{Rieke}, G.~H., {Young}, E.~T., {Engelbracht}, C.~W., {et~al.} 2004, \apjs,
  154, 25

\bibitem[{{Su{\'a}rez Mascare{\~n}o} {et~al.}(2016){Su{\'a}rez Mascare{\~n}o},
  {Rebolo}, \& {Gonz{\'a}lez Hern{\'a}ndez}}]{SMea16}
{Su{\'a}rez Mascare{\~n}o}, A., {Rebolo}, R., \& {Gonz{\'a}lez Hern{\'a}ndez},
  J.~I. 2016, \aap, 595, A12

\bibitem[{{Thuillier} {et~al.}(2004){Thuillier}, {Floyd}, {Woods}, {Cebula},
  {Hilsenrath}, {Hers{\'e}}, \& {Labs}}]{Tea04}
{Thuillier}, G., {Floyd}, L., {Woods}, T.~N., {et~al.} 2004, Advances in Space
  Research, 34, 256

\bibitem[{{Turbet} {et~al.}(2016){Turbet}, {Leconte}, {Selsis}, {Bolmont},
  {Forget}, {Ribas}, {Raymond}, \& {Anglada-Escud{\'e}}}]{Tea16}
{Turbet}, M., {Leconte}, J., {Selsis}, F., {et~al.} 2016, \aap, 596, A112

\bibitem[{{Veeder}(1974)}]{V74}
{Veeder}, G.~J. 1974, \aj, 79, 1056

\bibitem[{{Walker}(1981)}]{W81}
{Walker}, A.~R. 1981, \mnras, 195, 1029

\bibitem[{{Wargelin} {et~al.}(2017){Wargelin}, {Saar}, {Pojma{\'n}ski},
  {Drake}, \& {Kashyap}}]{Wea17}
{Wargelin}, B.~J., {Saar}, S.~H., {Pojma{\'n}ski}, G., {Drake}, J.~J., \&
  {Kashyap}, V.~L. 2017, \mnras, 464, 3281

\bibitem[{{Wood} {et~al.}(2005){Wood}, {Redfield}, {Linsky}, {M{\"u}ller}, \&
  {Zank}}]{Woodetal2005}
{Wood}, B.~E., {Redfield}, S., {Linsky}, J.~L., {M{\"u}ller}, H.-R., \& {Zank},
  G.~P. 2005, \apjs, 159, 118

\bibitem[{{Woodgate} {et~al.}(1998){Woodgate}, {Kimble}, {Bowers}, {Kraemer},
  {Kaiser}, {Danks}, {Grady}, {Loiacono}, {Brumfield}, {Feinberg}, {Gull},
  {Heap}, {Maran}, {Lindler}, {Hood}, {Meyer}, {Vanhouten}, {Argabright},
  {Franka}, {Bybee}, {Dorn}, {Bottema}, {Woodruff}, {Michika}, {Sullivan},
  {Hetlinger}, {Ludtke}, {Stocker}, {Delamere}, {Rose}, {Becker}, {Garner},
  {Timothy}, {Blouke}, {Joseph}, {Hartig}, {Green}, {Jenkins}, {Linsky},
  {Hutchings}, {Moos}, {Boggess}, {Roesler}, \& {Weistrop}}]{Wea98}
{Woodgate}, B.~E., {Kimble}, R.~A., {Bowers}, C.~W., {et~al.} 1998, \pasp, 110,
  1183

\bibitem[{{Wright} {et~al.}(2010){Wright}, {Eisenhardt}, {Mainzer}, {Ressler},
  {Cutri}, {Jarrett}, {Kirkpatrick}, {Padgett}, {McMillan}, {Skrutskie},
  {Stanford}, {Cohen}, {Walker}, {Mather}, {Leisawitz}, {Gautier}, {McLean},
  {Benford}, {Lonsdale}, {Blain}, {Mendez}, {Irace}, {Duval}, {Liu}, {Royer},
  {Heinrichsen}, {Howard}, {Shannon}, {Kendall}, {Walsh}, {Larsen}, {Cardon},
  {Schick}, {Schwalm}, {Abid}, {Fabinsky}, {Naes}, \& {Tsai}}]{Wea10}
{Wright}, E.~L., {Eisenhardt}, P.~R.~M., {Mainzer}, A.~K., {et~al.} 2010, \aj,
  140, 1868

\bibitem[{{Wright} \& {Drake}(2016)}]{WD16}
{Wright}, N.~J. \& {Drake}, J.~J. 2016, \nat, 535, 526

\bibitem[{{Wright} {et~al.}(2011){Wright}, {Drake}, {Mamajek}, \&
  {Henry}}]{Wea11}
{Wright}, N.~J., {Drake}, J.~J., {Mamajek}, E.~E., \& {Henry}, G.~W. 2011,
  \apj, 743, 48

\end{thebibliography}

\end{document}